\ifx\allfiles\undefined
\documentclass[a4paper,fleqn]{cas-sc}
\usepackage{rotating}
\usepackage{appendix}
\usepackage{graphicx}
\usepackage{caption}
\usepackage{tabularray}
\usepackage[authoryear]{natbib}
\usepackage{float}
\usepackage{tabularx}
\usepackage{makecell}
\usepackage{amsmath}
\usepackage{mathtools}
\usepackage{color}
\usepackage{booktabs}
\usepackage{array}
\usepackage{pdflscape}
\usepackage{longtable}
\usepackage{epstopdf}
\usepackage{grfext}
\usepackage{pdfpages}
\usepackage{microtype}
\usepackage{xcolor}

\def\tsc#1{\csdef{#1}{\textsc{\lowercase{#1}}\xspace}}
\tsc{WGM}
\tsc{QE}
\begin{document}
\let\WriteBookmarks\relax
\def\floatpagepagefraction{1}
\def\textpagefraction{. 001}

\shorttitle{} 

\shortauthors{ {}}

\title [mode = title]{
%Population-based accessibility and double disadvantage in health-supportive urban services: Evidence from Hefei, China
The Moving Target of Urban Equity: Spatiotemporal Demand and Double Disadvantage in Hefei, China}
% Author Information
\author[label1,label4]{Shirui Zhou}
% \ead{shiruizhou@tju.edu.cn}

\author[label3]{Matteo Bruno}
\cormark[1]
\ead{matteo.bruno@sony.com}

\author[label6]{Mattia Mazzoli}

\author[label1,label2]{Junfang Tian}
\cormark[1]
\ead{jftian@tju.edu.cn}

\author[label5]{Rui Jiang}

\author[label7]{Enwan Zhang}

\author[label7]{Zheng Li}

\author[label4,label3,label8]{Vittorio Loreto}

\affiliation[label1]{organization={Institute of Systems Engineering, College of Management and Economics, Tianjin University},
addressline={No. 92 Weijin Road},
city={Nankai District},
postcode={300072},
state={Tianjin},
country={China}}

\affiliation[label4]{organization={Physics Department, Sapienza University of Rome},
addressline={Piazzale Aldo Moro, 5},
city={Rome},
state={RM},
postcode={00185},
country={Italy}}

\affiliation[label3]{organization={Sony Computer Science Laboratories – Rome, Joint Initiative CREF-SONY, Centro Ricerche Enrico Fermi},
addressline={Via Panisperna, 89/a},
city={Rome},
postcode={00184},
state={RM},
country={Italy}}

\affiliation[label6]{organization={ISI Foundation},
addressline={via Della Rocca 20},
postcode={10123},
state={Turin},
country={Italy}}

\affiliation[label2]{organization={Laboratory of Computation and Analytics of Complex Management Systems (CACMS),Tianjin University},
addressline={No. 92 Weijin Road},
city={Nankai District},
postcode={300072},
state={Tianjin},
country={China}}

\affiliation[label5]{organization={School of Systems Science, Beijing Jiaotong University},
addressline={No. 3 Shangyuancun},
city={Haidian District},
postcode={100044},
state={Beijing},
country={China}}

\affiliation[label7]{organization={China Mobile Communications Group Anhui Co., Ltd},
addressline={No. 99, Changjiang West Road},
postcode={230031},
state={Anhui},
country={China}}

\affiliation[label8]{organization={Complexity Science Hub Vienna},
addressline={Metternichgasse 8},
postcode={1030},
state={Vienna},
country={Austria}}

\cortext[1]{Corresponding author. }

% \nonumnote{}

\begin{abstract}
Equitable access to essential urban services is a pillar of modern planning, yet most accessibility models rely strictly on static residential locations, ignoring how demand shifts throughout the daily loop. This study introduces a population-based, temporally differentiated framework to examine the resulting "moving target" of urban equity, focusing on medical facilities and green spaces in Hefei, China. Utilising large-scale mobile phone GPS data, we construct dynamic residential and workplace population exposure surfaces to capture shifting hourly demand. We then evaluate accessibility via network-based travel times paired with a novel per-capita provision metric that accounts for real-time demand competition. We define \textit{double disadvantage} as the co-occurrence of poor spatial accessibility and insufficient per-capita service availability. Counterintuitively, the results reveal that double-disadvantaged areas cluster primarily along the inner suburban belt rather than the remote periphery, where per-capita service provision remains relatively sufficient. Furthermore, temporal shifts drastically alter equity landscapes: daytime workplace concentrations intensely exacerbate demand competition in urban job centres. These findings demonstrate that urban inequality depends heavily on spatiotemporal population flows rather than just the fixed location of services. Ultimately, achieving true urban equity requires dynamic planning interventions that address time-varying demand rather than focusing solely on static, home-based metrics.
\end{abstract}

% Keywords

\maketitle
% Main text

\section{Introduction}

Accessibility has long been regarded as a central concept in transport research and urban planning, particularly in discussions of social equity and transport justice \citep{hansen1959accessibility,lucasTransportSocialExclusion2012,martens2016transport,levinson2020towards}. Rather than focusing solely on mobility or travel efficiency, accessibility reflects individuals' ability to reach essential opportunities such as healthcare, green spaces, education, and daily services through the interaction of transport systems, land-use patterns, and population distribution \citep{bertolini1996nodes,geursAccessibilityEvaluationLanduse2004a,bruno2026dimensions}. Within this perspective, improving accessibility through low-cost and health-promoting transport modes, especially walking and cycling, has become a central objective of contemporary urban policy aimed at reducing urban emissions \citep{marzolla2026proximity,marzolla2026compact} and social and spatial inequalities \citep{paezMeasuringAccessibilityPositive2012a,neutensAccessibilityEquityHealth2015a,valeAccessibilityInequalityEurope2023}.

The 15-minute city concept has recently emerged as a prominent planning framework that operationalises accessibility in explicitly time-based terms \citep{morenoIntroducing15MinuteCity2021a,allam15minuteCityOffersNew2022,calafiore20minuteCityEquityAnalysis2022a}. By emphasising that essential daily services should be reachable within a 15-minute travel time using active or non-motorised modes, the framework reframes urban equity as a matter of everyday accessibility rather than static spatial proximity \citep{ellder15minuteCityDilemmaBalancing2024}. The concept has gained substantial policy traction worldwide \citep{valeAccessibilityInequalityEurope2023,weng15minuteWalkableNeighborhoods2019} and aligns with broader agendas related to health, sustainability, and inclusive urban development \citep{colglazierSustainableDevelopmentAgenda2015}. In China, this approach has been institutionalised through initiatives such as the “15-minute living circle” and the “15-minute health service circle” \citep{HealthyChina2030_NHC_2016,StateCouncil_14thFYP_NationalHealth_2022}, which explicitly prioritise equitable access to health-supportive urban services through walking and cycling.

Despite its growing policy relevance, empirical research on accessibility equity within the 15-minute city paradigm remains conceptually limited in several important respects. Most existing studies evaluate accessibility primarily from the perspective of opportunity distribution, often relying on points-of-interest (POI) data to assess whether services are located within a given travel-time threshold \citep{profiroiu15minuteChallengeEvaluatingGaps2025,chenCompactUrbanMorphology15minute2025,wang15minuteCityUrbanCore2025}. Although recent contributions increasingly incorporate network-based travel times and realistic transport modes, accessibility is still commonly treated as a static property of places rather than as an outcome shaped by the interaction between population distribution, daily mobility, and service availability \citep{fanShenzhens15minuteCityInitiative2025,popescuHowSmart15MinuteCity2025,luDeveloping15minuteCityPolicy2025}. Consequently, many evaluations emphasize where services are located while paying limited attention to how individuals actually encounter these opportunities through their daily activity patterns \citep{rahmanImpactsPointInterestPOI2025}. A universal measurement framework for the 15-minute city has recently been proposed by \citet{brunoUniversalFrameworkInclusive15minute2024}, again evaluating accessibility within time-based planning paradigms but considering only the residential population distribution in the evaluation of proximity.
\begin{figure}[H]
\centering
\includegraphics[width=0.95\linewidth]{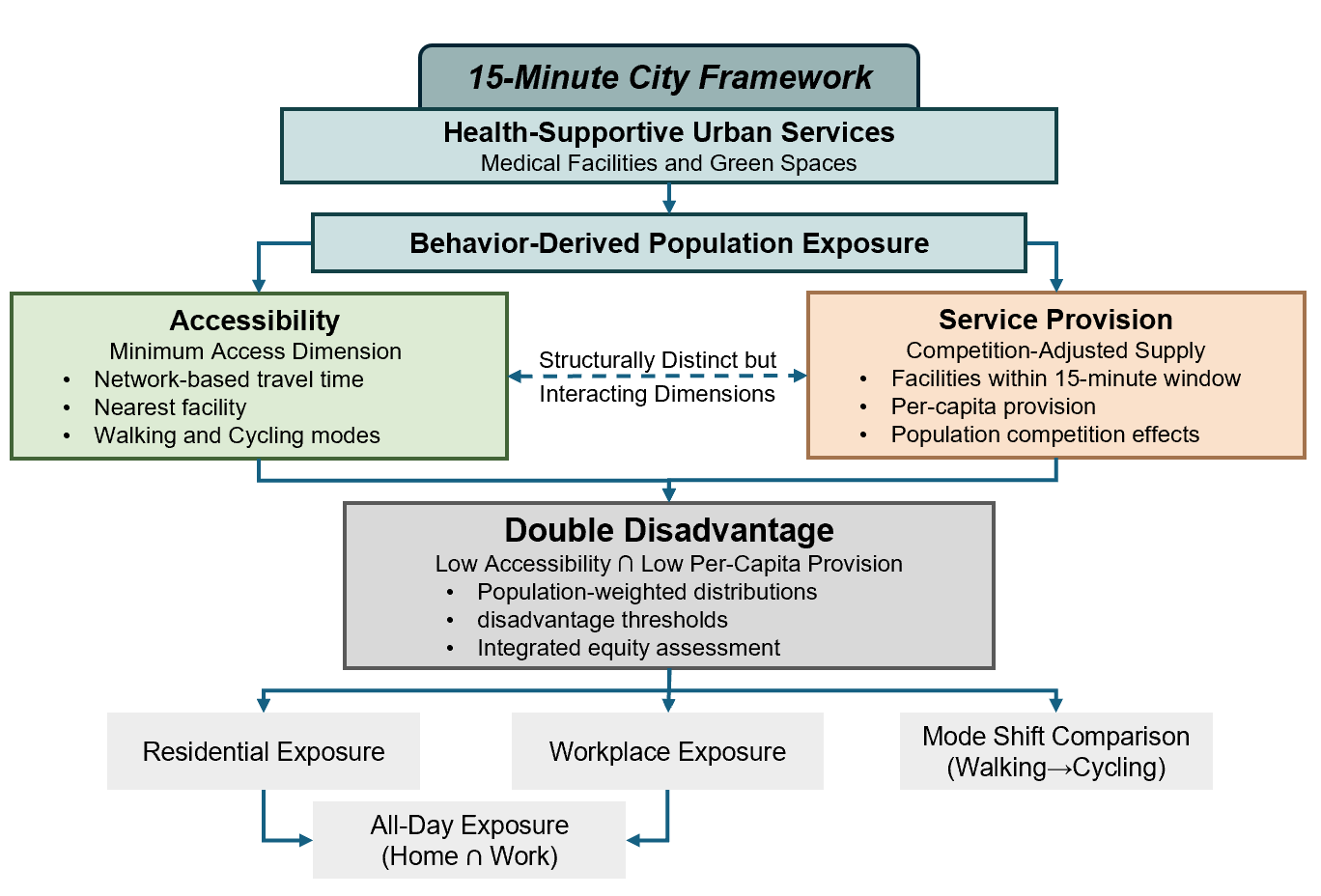}
\caption{Conceptual framework of the population-based double-disadvantage assessment. Accessibility (minimum travel time) and competition-adjusted service provision (per-capita supply within a 15-minute window) are treated as structurally distinct but interacting dimensions under dynamic population exposure (residential and workplace contexts). Double disadvantage emerges where low relative position in both dimensions persists across exposure contexts and transport modes.}
\label{fig:framework}
\end{figure}

A first limitation of existing research concerns the role of population competition in shaping effective access to services. In dense or rapidly urbanising environments, favourable travel-time accessibility does not necessarily translate into adequate service availability at the individual level. High population density can substantially dilute per-capita access to facilities, creating situations in which areas with good nominal accessibility nevertheless experience significant service pressure. Previous studies have therefore emphasised the importance of competition-adjusted accessibility indicators \citep{goddardEquityAccessHealth2001,luoMeasuresSpatialAccessibility2003,apparicioComparingAlternativeApproaches2008,mcgrailMeasuringSpatialAccessibility2009,neutensEquityUrbanService2010}. However, accessibility and service provision are still frequently examined in isolation, limiting the ability to identify locations where poor physical access and insufficient service capacity coexist and jointly reinforce disadvantage.

A second limitation concerns how population exposure is conceptualised in accessibility assessments. Most studies implicitly anchor exposure to public services at individuals' residential locations \citep{kwanSpaceWeKnew2013}. Yet daily mobility, especially commuting between home and workplace, systematically redistributes population across space and time \citep{neutensAnalysisDaytodayVariations2012}. Daytime population concentrations in employment centres may intensify competition for services, while residential neighbourhoods may experience different accessibility conditions during non-working hours \citep{ryanWhatAreWe2021,zhangAssessingInequalityXminuteCity2025a,zhangUrbanFoodDeliveryServices2025}. Empirical evidence based on mobile phone, GPS, and other location-based data has demonstrated substantial discrepancies between residential and daytime population distributions \citep{abbiasov15minuteCityQuantifiedUsing2024}, suggesting that residence-based accessibility assessments may misrepresent both the intensity and spatial distribution of service competition.

These limitations are particularly consequential in the context of the 15-minute city, where accessibility is often evaluated using static snapshots of opportunity coverage \citep{ermagun15minuteCityDichotomyTimedenominated2025}. By focusing primarily on the spatial availability of services rather than on dynamic population exposure, existing approaches risk overstating effective access and overlooking forms of inequality that emerge from temporal variation and population competition \citep{valeEffectiveAccessibilityUsing2020,calafiore20minuteCityEquityAnalysis2022a}. Addressing these challenges requires a shift from opportunity-centred evaluations toward a population-based perspective that explicitly considers how accessibility and service provision are experienced across different activity contexts \citep{ellder15minuteCityDilemmaBalancing2024}.

Against this background, this study focuses on health-supportive urban services, specifically medical facilities and green spaces \citep{ekkelNearbyGreenSpace2017}. These services play a critical role in the 15-minute city agenda and are closely linked to both physical \citep{Wu2026GreennessHypertension} and mental well-being \citep{twohig-bennettHealthBenefitsGreat2018}. At the same time, they exhibit distinct spatial and functional characteristics that make them particularly sensitive to interactions among accessibility, population competition, and daily exposure. Medical services often involve centralised provision structures and strong demand concentration, whereas access to green spaces tends to be more spatially dispersed and embedded within everyday activity environments \citep{fanAccessibilityPublicUrban2017}. Examining these two facility types, therefore, provides a useful basis for understanding how different forms of service provision interact with mobility conditions and population exposure \citep{nieuwenhuijsenEvaluation330300Green2022,benati2024unequal,wang15minuteCityUrbanCore2025}.

%Also add something on the work accessibility with citations (proximity-based, the dimensions of, cities beyond?)

Building on this motivation, this study develops a population-based, activity-oriented framework to evaluate equity in access to health-supportive urban services within the 15-minute city paradigm. Leveraging large-scale mobile phone GPS data, we identify individual residential and workplace locations and construct dynamic population distributions that capture temporal patterns of daily activity rather than static residential distributions. Accessibility to medical services and green spaces is assessed under walking and cycling modes using network-based travel times, while service provision is measured through competition-adjusted per-capita indicators within a 15-minute travel-time window.

As illustrated in Fig.~\ref{fig:framework}, the proposed framework conceptualises service equity as the interaction between three structural components: spatial reach (minimum travel time), competition intensity (per-capita service provision within the accessibility window), and temporal redistribution of population exposure across residential and workplace contexts. Double disadvantage arises where low relative positions in both accessibility and service provision persist across exposure contexts, providing an integrated basis for identifying structurally persistent inequities.

This study makes two main contributions. First, it extends accessibility and equity research beyond conventional residence-based, opportunity-oriented assessments by incorporating behaviour-derived population exposure from large-scale mobility data and explicitly comparing residential and workplace contexts. In doing so, it shows how daily population redistribution reshapes demand pressure and reveals important differences in effective access to urban services between home and work. Second, it develops a unified analytical framework that combines physical accessibility with competition-adjusted service provision and introduces the concept of \textit{double disadvantage} to identify communities where poor access and insufficient service capacity jointly reinforce inequality. Together, these contributions demonstrate that accessibility disparities cannot be understood solely from travel time, but must be examined through the interaction among population exposure, service capacity, and competition.

\section{Data and Methods}

\subsection{Data Description}

This study focuses on Hefei, the capital city of Anhui Province, China. As of the end of 2024, Hefei has a permanent population of approximately 10 million and comprises nine county-level administrative units. The city exhibits a polycentric spatial structure, characterised by a dominant central urban area complemented by multiple secondary centres located in surrounding districts and counties. In recent years, Hefei has experienced rapid urban expansion accompanied by the development of several new urban districts, making it a representative case for examining accessibility and equity issues in the context of contemporary urban growth.

Hefei has actively promoted people-centred urban development policies, including the implementation of the “15-minute living circle” framework, which emphasises access to essential daily services through walking and cycling. The coexistence of a dense central area, newly developed urban districts, and peripheral county-level centres provides a heterogeneous urban environment in which accessibility conditions and service provision vary substantially across space.

\subsubsection{Mobile Phone GPS Data}

The analysis is based on large-scale mobile phone GPS data collected from smartphone applications with user consent, covering a continuous two-month period from October to November 2024. The dataset includes approximately 5.38 million anonymised users within Hefei, representing the full population of China Mobile app users in the city. Given that China Mobile accounts for roughly 60\% of the national mobile market, the dataset provides extensive population coverage and captures a substantial share of daily urban mobility.

The GPS records provide meter-level spatial accuracy, enabling precise identification of individual activity locations. The temporal frequency of observations varies across users and depends on application usage behaviour, leading to irregular, relatively sparse location traces compared with traditional signalling data. Nevertheless, the high spatial precision of GPS data makes it well-suited for detecting recurrent activity locations, such as residential and workplace locations, as well as frequently visited non-work and non-home activity locations.

All data were anonymised prior to analysis, and no personal identifiers were available, ensuring that individual users could not be re-identified. The use of aggregated indicators in subsequent analyses further protects user privacy.

\subsubsection{Points of Interest Data}

Data on public service facilities were obtained through the Amap (Gaode) application programming interface (API) in 2025. The POI dataset adopts a three-level hierarchical classification system, allowing facilities to be consistently categorised across the study area. This study focuses on two types of health-supportive urban opportunities: medical facilities and green spaces.

Medical facilities include POIs classified under the following mid-level categories: comprehensive hospitals, specialised hospitals, clinics, emergency centres, and disease prevention institutions. Green spaces include POIs classified as scenic areas, parks and plazas, zoos, botanical gardens, leisure spaces, and resort or recuperation areas. These categories represent publicly accessible environments that contribute to physical and mental well-being in everyday urban life.

\subsubsection{Transport Network and Travel Time Estimation}

Walking and cycling travel times were estimated using transport networks derived from OpenStreetMap (OSM). Separate network representations were constructed for walking and cycling, in accordance with standard OSM specifications. Travel times were calculated based on network distances and default non-motorised travel speeds commonly adopted in OSM-based analyses. No additional assumptions regarding congestion, signal delays, or individual route choice behaviour were introduced, allowing the estimated travel times to reflect generalised accessibility conditions for non-motorised transport.

\subsubsection{Spatial Units and Data Preprocessing}

All spatial analyses were conducted using the H3 hexagonal grid system to ensure consistent spatial representation across datasets. H3 resolution 9 was adopted, corresponding to an average edge length of approximately 0.201 km and a cell area of about 0.105 km$^2$. 
To avoid spurious accessibility assessments, the analysis was restricted to land-based grid cells with identified residential populations. Water bodies were excluded using land--water masks derived from the Natural Earth dataset. This preprocessing step ensures that accessibility and equity indicators are computed only for inhabited areas and are not affected by uninhabitable spatial units.

\subsection{Methods}

\subsubsection{Population spatialization and activity locations}

Large-scale mobile phone GPS data were used to infer individual residential and workplace locations from recurrent spatiotemporal activity patterns. A key advantage of these data is that they allow residential and workplace locations to be linked for the same individuals, making it possible to estimate home-work differences in accessibility at the individual level and at high spatial resolution. By contrast, census data typically provide population aggregates or commuting origin--destination flows at the tract or district level, which are less suited to capturing fine-grained, person-specific differences in spatial exposure. The identified home and workplace locations were therefore aggregated using the H3 hexagonal grid system to construct fine-grained population exposure surfaces.

From the original dataset of approximately 5.38 million anonymised users, residential and workplace locations could not be identified for all individuals because app-based GPS records are irregular and usage-dependent. In addition, the raw data may include short-term visitors or cross-city users whose activity traces do not represent stable residence–-workplace routines within Hefei. To ensure strict comparability across residential, workplace, and all-day exposure analyses, we therefore restricted the analytical sample to users for whom both home and workplace locations could be reliably identified within the study area. This procedure helps exclude transient populations and ensures that differences across exposure contexts reflect within-individual changes in spatial exposure rather than compositional differences introduced by temporary visitors.

This procedure retained 1{,}330{,}384 users for subsequent analyses. Each retained individual contributes one residential location and one workplace location to the corresponding population exposure surfaces. Consequently, observed differences between residential- and workplace-based accessibility or service provision reflect differences in spatial exposure for the same underlying population, rather than compositional differences between separate samples.

\subsubsection{Accessibility measures}

Accessibility was measured at the H3 grid level for walking and cycling using network-based travel times derived from OpenStreetMap. For each grid cell, the geometric centroid of the hexagon was used as the origin point for travel time calculations.

Accessibility to public services was operationalised as the average travel time to the 20 nearest facilities of a given type. This procedure is the same as the Proximity Time introduced in \citet{brunoUniversalFrameworkInclusive15minute2024} and captures a basic dimension of physical access to essential services in everyday urban environments. Because this study is situated within the 15-minute city framework, which prioritises access on foot and by bike, we focused exclusively on walking and cycling rather than motorised transport. Accessibility was calculated separately for medical facilities and green spaces using shortest-path travel times on mode-specific transport networks.

% This approach provides a consistent and interpretable measure of potential accessibility based on network travel time, which is appropriate for the analytical objective of this study. In particular, it allows us to isolate the role of physical separation between population locations and service facilities, and to combine this travel-time-based access dimension with competition-adjusted service provision in identifying compounded disadvantage. 

\subsubsection{Competition-adjusted per-capita service provision}

To account for population demand and competition effects, service provision was measured using per-capita exposure indicators. For each H3 grid cell $i$, per-capita provision was defined as
\begin{equation}
P_i(t)=\frac{N_i(t)}{\tilde{Pop}_i},
\end{equation}

where $N_i(t)$ denotes the number of facilities of a given type, such as green spaces or medical facilities, that are accessible from grid cell $i$ within a travel-time threshold $t$, and $\tilde{Pop}_i$ denotes the estimated residential population when the residential exposure context is analysed and the estimated workplace population when the workplace exposure context is analysed. In this study, $t$ is fixed at 15 minutes.

\begin{figure}[H]
\centering
\includegraphics[width=0.4\textwidth]{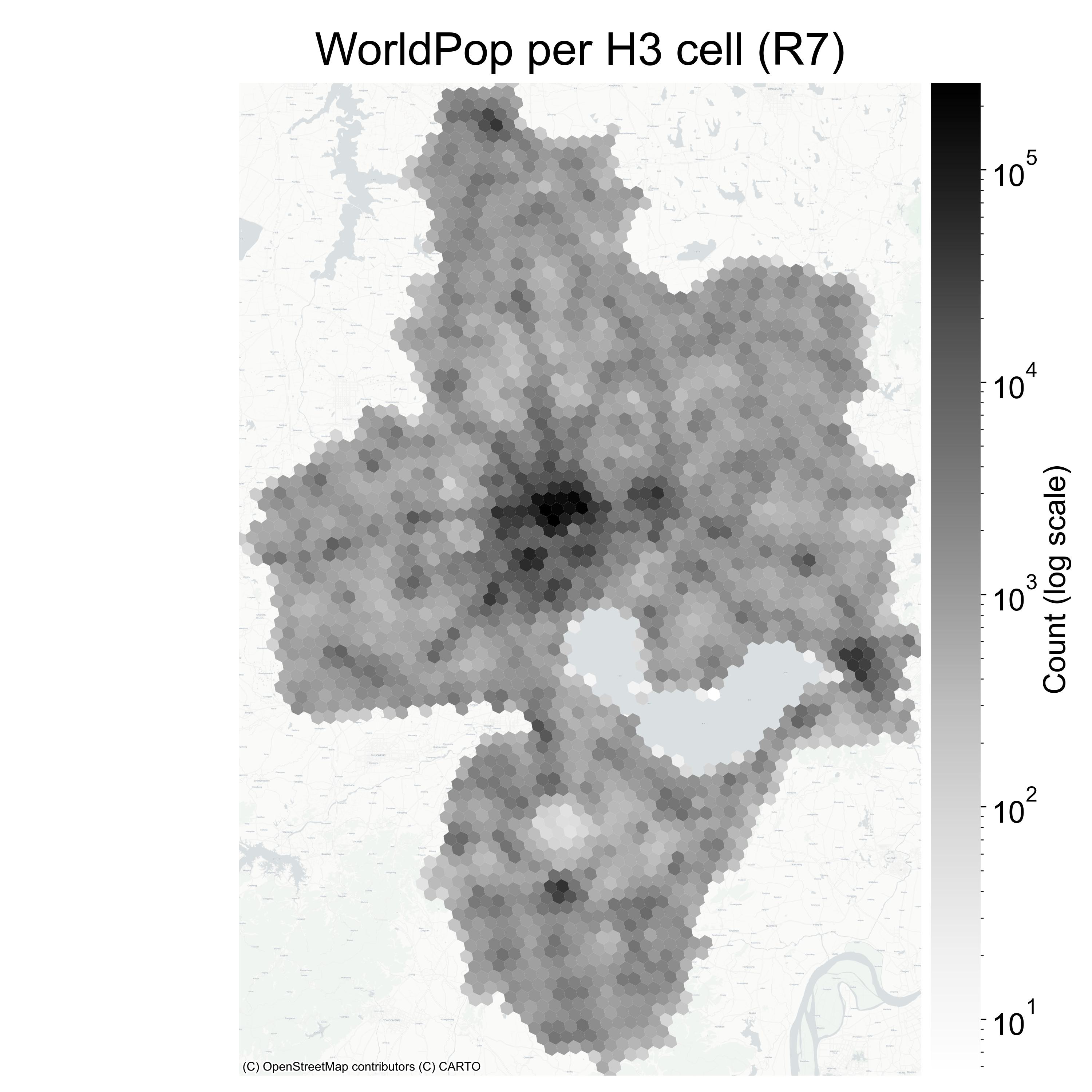}
\caption{Population distributions from WorldPop (\url{https://www.worldpop.org/}).}
\label{fig:worldpop_data}
\end{figure}

\begin{figure}[H]
\centering
\includegraphics[width=0.6\linewidth]{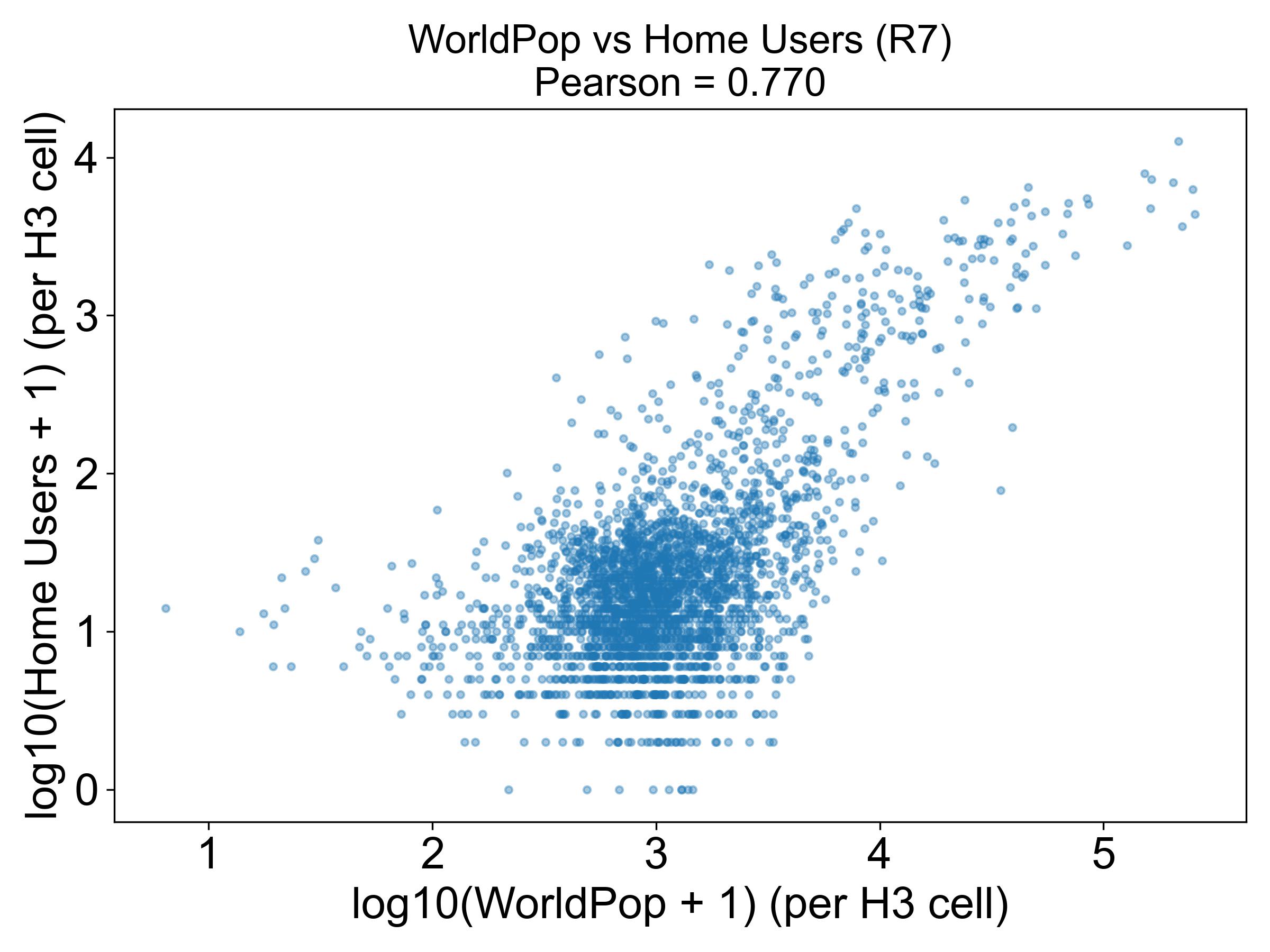}
\caption{Comparison between behaviour-derived residential population and WorldPop gridded population data at the H3 level.
A strong spatial correspondence is observed, while deviations highlight differences between static census-based and behaviour-derived population distributions.}
\label{fig:worldpop_comparison}
\end{figure}

The behaviour-derived population counts obtained from mobile phone GPS records constitute a statistical sample rather than the full population of Hefei. Specifically, after identifying home and workplace locations from recurrent spatiotemporal activity patterns, we aggregated the number of sampled users assigned to each H3 grid cell across exposure contexts. Thus, $Pop_{i,c}$ denotes the observed number of sampled users in grid cell $i$ under context $c$, where $c \in \{\mathrm{home}, \mathrm{work}\}$. Let $N_{\text{sample},c}=\sum_i Pop_{i,c}$ denote the total sampled population under context $c$. To express per-capita provision relative to the official city population, we rescaled the population of each grid cell as
\begin{equation}
\tilde{Pop}_{i,c}=Pop_{i,c}\times \frac{N_{\text{Hefei}}}{N_{\text{sample},c}},
\end{equation}
where $N_{\text{Hefei}}$ is the official total population of Hefei. This proportional scaling aligns the aggregate sample population with the official city total while preserving the relative spatial distribution observed in the GPS-based sample within each exposure context. We acknowledge, however, that this procedure assumes a broadly stable spatial penetration rate of the GPS sample across the city. If sample penetration varies systematically across locations, the rescaled population surfaces may inherit this bias. To assess this risk, we compared the behaviour-derived residential population with independent WorldPop gridded population data at the H3 level (Appendix A, Figs.~\ref{fig:worldpop_data} and~\ref{fig:worldpop_comparison}), which shows a strong positive spatial correspondence (Pearson $=0.770$). This comparison does not eliminate the possibility of local penetration bias, but it suggests that the overall spatial pattern of the sample population is broadly consistent with an external population benchmark.

% We adopt A 15-minute travel-time threshold for both walking and cycling modes to reflect the effective service environment. We adopt this threshold for consistently with the policy logic of the 15-minute city framework, commonly used in planning practice, rather than assuming continuous decay of service provision with time.

Per-capita provision was computed separately for residential and workplace populations to capture context-dependent exposure, extending accessibility frameworks that focus mainly on residential locations. Unlike travel time, it captures population pressure within a fixed access window, so the disadvantage is identified through spatial differences in competition-adjusted exposure rather than raw time-based access.

\subsubsection{Definition of double disadvantage}

Building on the joint consideration of accessibility and per-capita provision, this study defines double disadvantage as the simultaneous occurrence of poor physical access and insufficient competition-adjusted service availability. For each combination of facility type, transport mode, and population context, disadvantage thresholds were defined using population-weighted distributions.

Specifically, grid cells were classified as having poor accessibility if their nearest-facility travel times fell within the worst-performing quartile of the population-weighted accessibility distribution. %To reduce the influence of right-skewness in travel-time distributions, quartile thresholds were computed based on the logarithmically transformed accessibility values, while retaining original travel-time values for interpretation and mapping. 
Similarly, insufficient service provision was defined as per-capita provision levels falling within the lowest quartile of the corresponding population-weighted distribution. Grid cells meeting both criteria were classified as \textbf{double disadvantaged}.

Thresholds were computed separately for residential and workplace populations to reflect differences in daytime and nighttime exposure and service competition, and for walking and cycling modes.

% Thresholds were computed separately for residential and workplace populations to account for systematic differences in exposure patterns and service competition between nighttime and daytime environments. This relative, distribution-based approach avoids reliance on arbitrary absolute cutoffs and enables consistent identification of compounded disadvantage across heterogeneous urban contexts.

% All analyses were conducted separately for walking and cycling modes, and for residential and workplace population surfaces. Accessibility, per-capita provision, population-weighted distributions, and double disadvantage were examined jointly to assess spatial patterns of service access, inequality, and the effects of transport mode shifts under the 15-minute city framework.

\begin{figure}[H]
\centering
\begin{minipage}[t]{0.47\textwidth}
    \textbf{(a)}\\[2pt]
    \includegraphics[width=\linewidth]{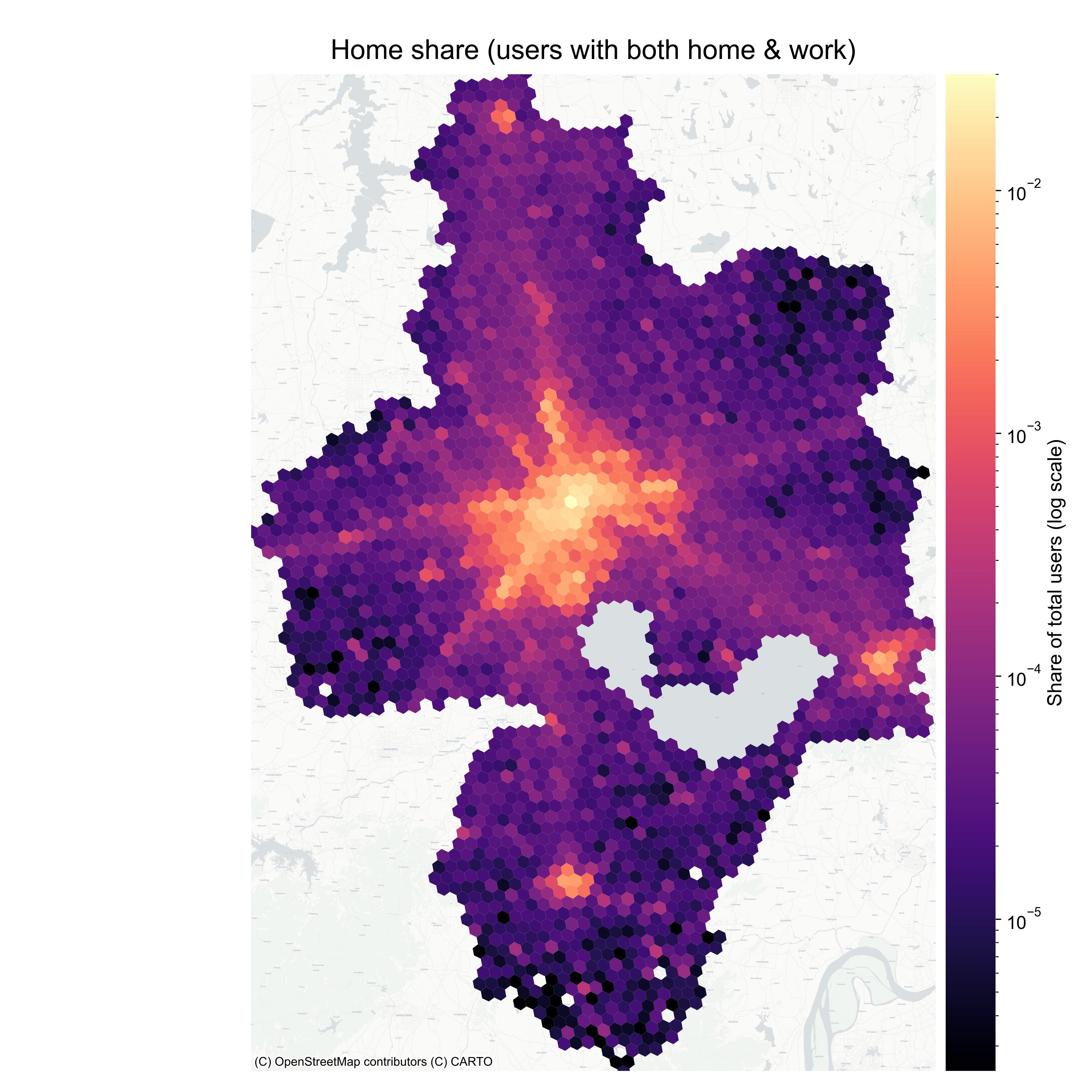}
\end{minipage}
\hspace{0.03\textwidth}
\begin{minipage}[t]{0.47\textwidth}
    \textbf{(b)}\\[2pt]
    \includegraphics[width=\linewidth]{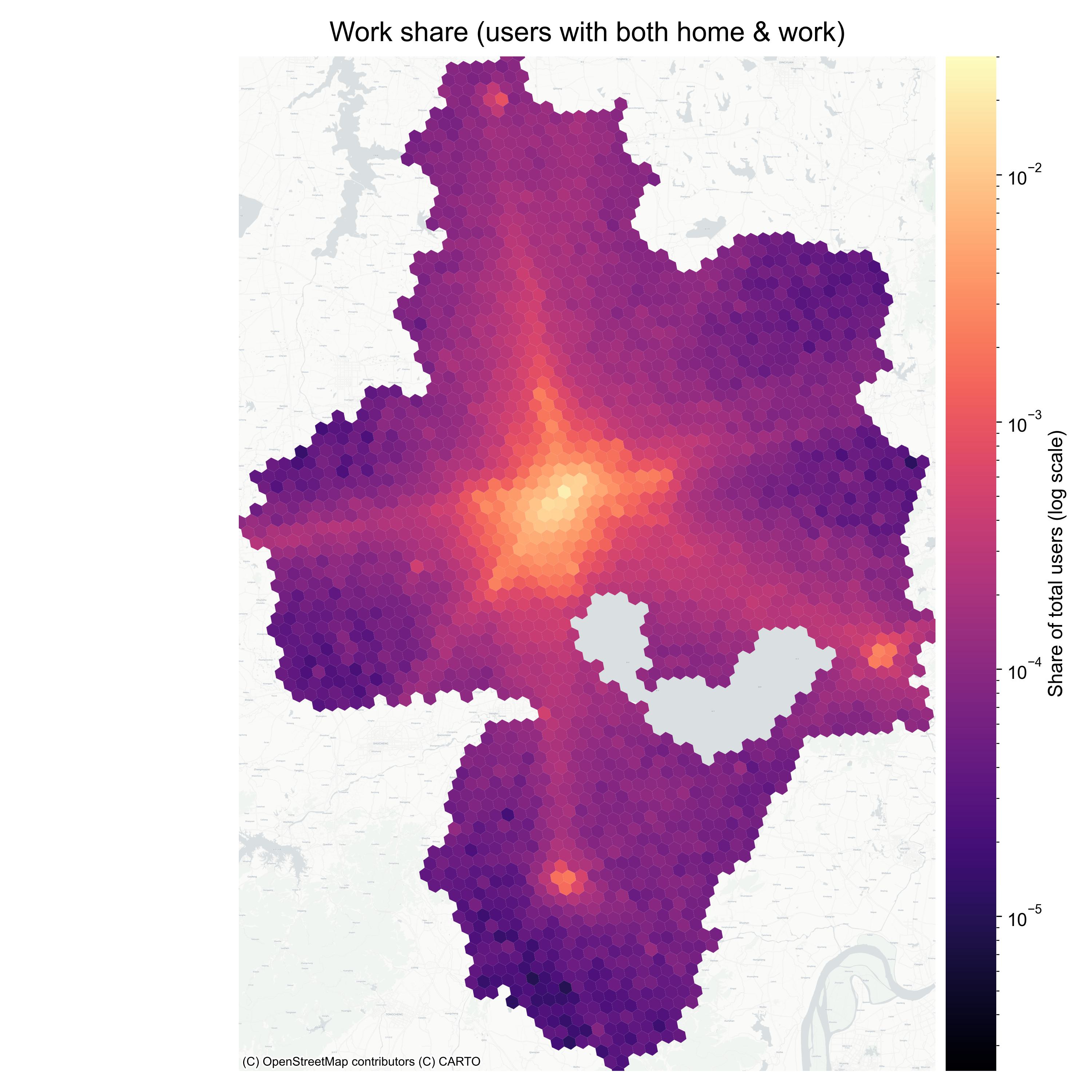}
\end{minipage}
\caption{
Normalised behaviour-derived population distributions aggregated to H3 grids.
(a) Residential population distribution inferred from nighttime GPS records.
(b) Workplace population distribution inferred from daytime GPS records.
Both panels use the same logarithmic colour scale, so colours are directly comparable as population shares rather than raw counts.
}
\label{fig:population_home_work}
\end{figure}

\section{Results}

\subsection{Behaviour-derived population redistribution}

Fig.~\ref{fig:population_home_work} presents behaviour-derived residential and workplace population distributions aggregated to H3 grids. Residential locations, inferred from nighttime GPS records, exhibit a relatively dispersed structure extending from the urban core toward peripheral districts and multiple subcentres. 

In contrast, workplace locations inferred from daytime GPS records are markedly more concentrated in the central urban area and along major roads and corridors, while peripheral zones exhibit lower workplace densities. The workplace surface, therefore, exhibits a steeper spatial gradient than the residential surface.

This redistribution between residential and working hours fundamentally reshapes exposure to urban services. Because workplace populations are more spatially concentrated, competition intensity in central areas is expected to increase during the daytime. All subsequent accessibility and provision analyses are therefore reported separately for residential and workplace populations.

\begin{figure}[H]
\centering
\begin{minipage}[t]{0.47\textwidth}
    \textbf{(a)}\\[2pt]
    \includegraphics[width=\linewidth]{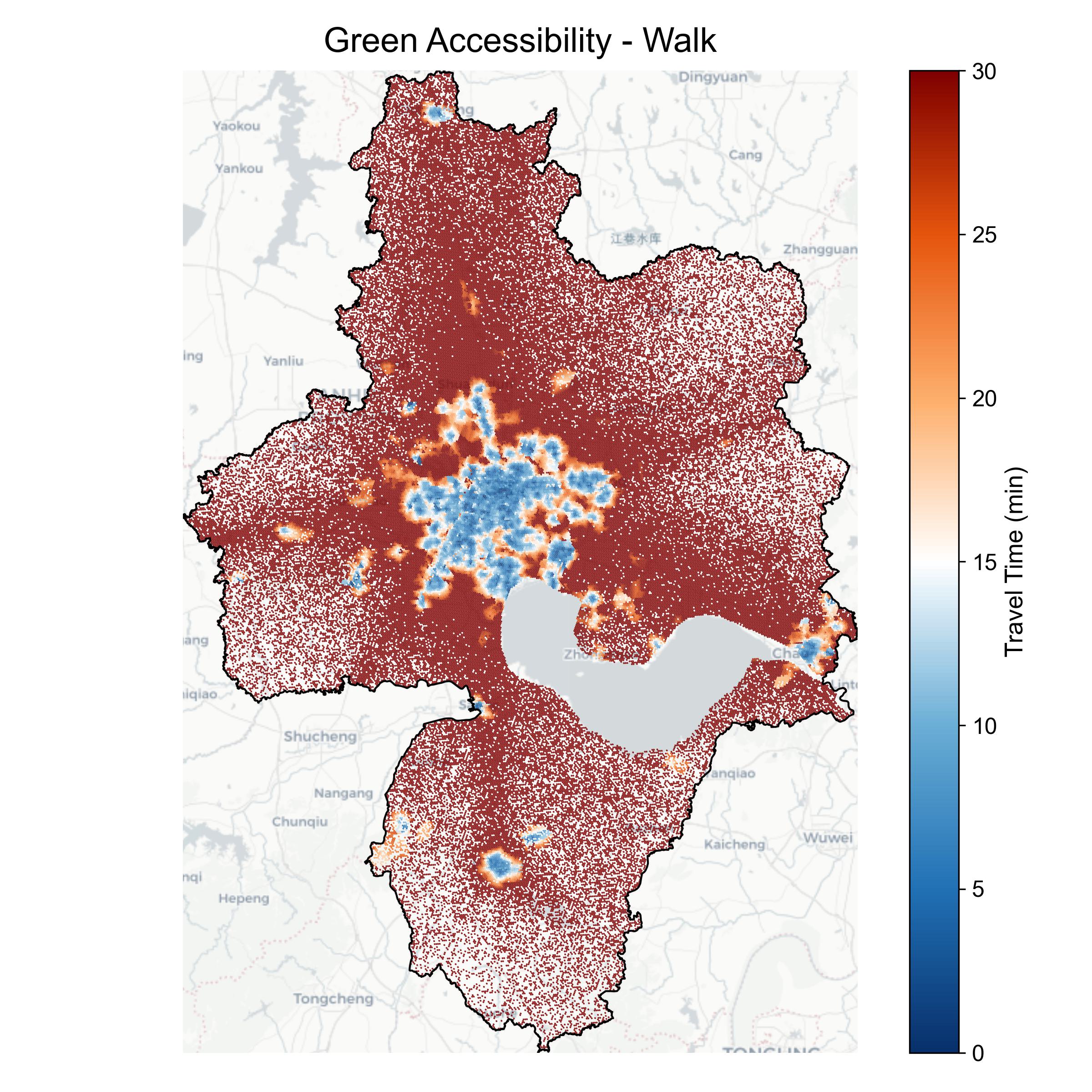}
\end{minipage}
\hspace{0.03\textwidth}
\begin{minipage}[t]{0.47\textwidth}
    \textbf{(b)}\\[2pt]
    \includegraphics[width=\linewidth]{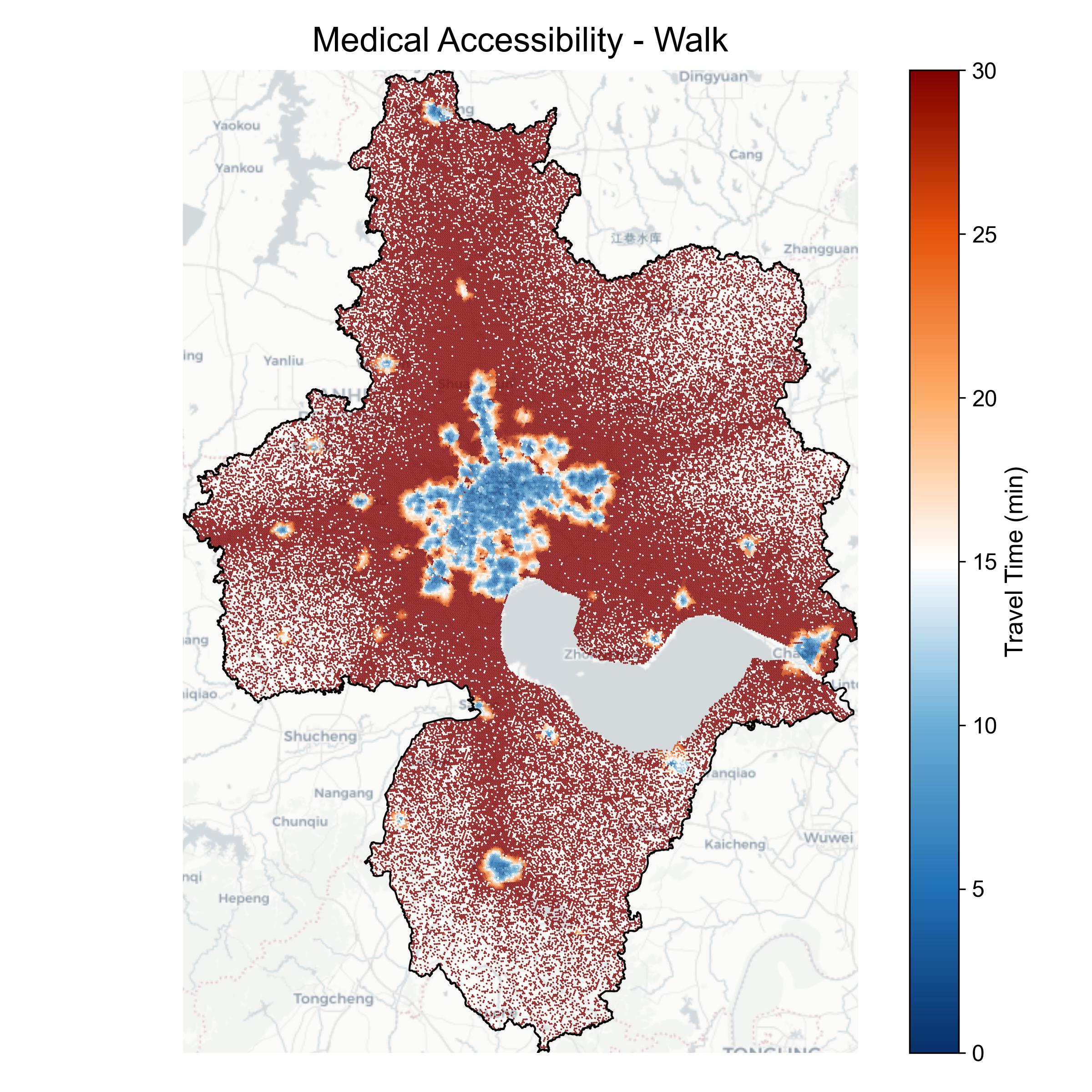}
\end{minipage}

\medskip

\begin{minipage}[t]{0.62\textwidth}
    \textbf{(c)}\\[2pt]
    \includegraphics[width=\linewidth]{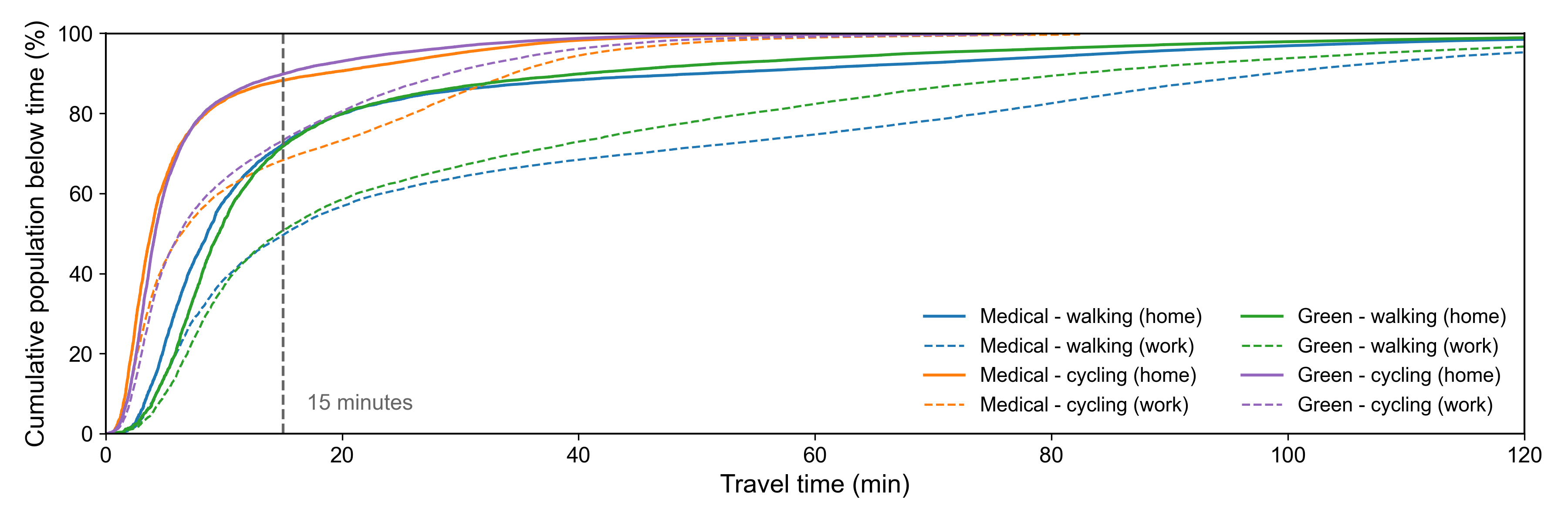}
\end{minipage}

\caption{Spatial distribution of H3-cell proximity time and population-weighted CDFs.
(a) Green space accessibility by walking.
(b) Medical service accessibility by walking.
Colours represent unweighted travel time for retained H3 cells; population counts were used only to mask cells with no observed home or work users.
The results of cycling are shown in the appendix (Fig.~\ref{fig:accessibility_maps_bike}).
(c) Population-weighted CDFs of proximity time for medical facilities and green spaces under walking and cycling, shown separately for residential and workplace populations.}
\label{fig:accessibility_maps}
\end{figure}

\subsection{Proximity time: the physical reach accessibility dimension}

We first examine nearest-facility travel time, which captures the physical reach of services independent of competition intensity.

\begin{figure}[H]
\centering
\begin{minipage}[t]{0.4\textwidth}
    \textbf{(a)}\\[2pt]
    \includegraphics[width=\linewidth]{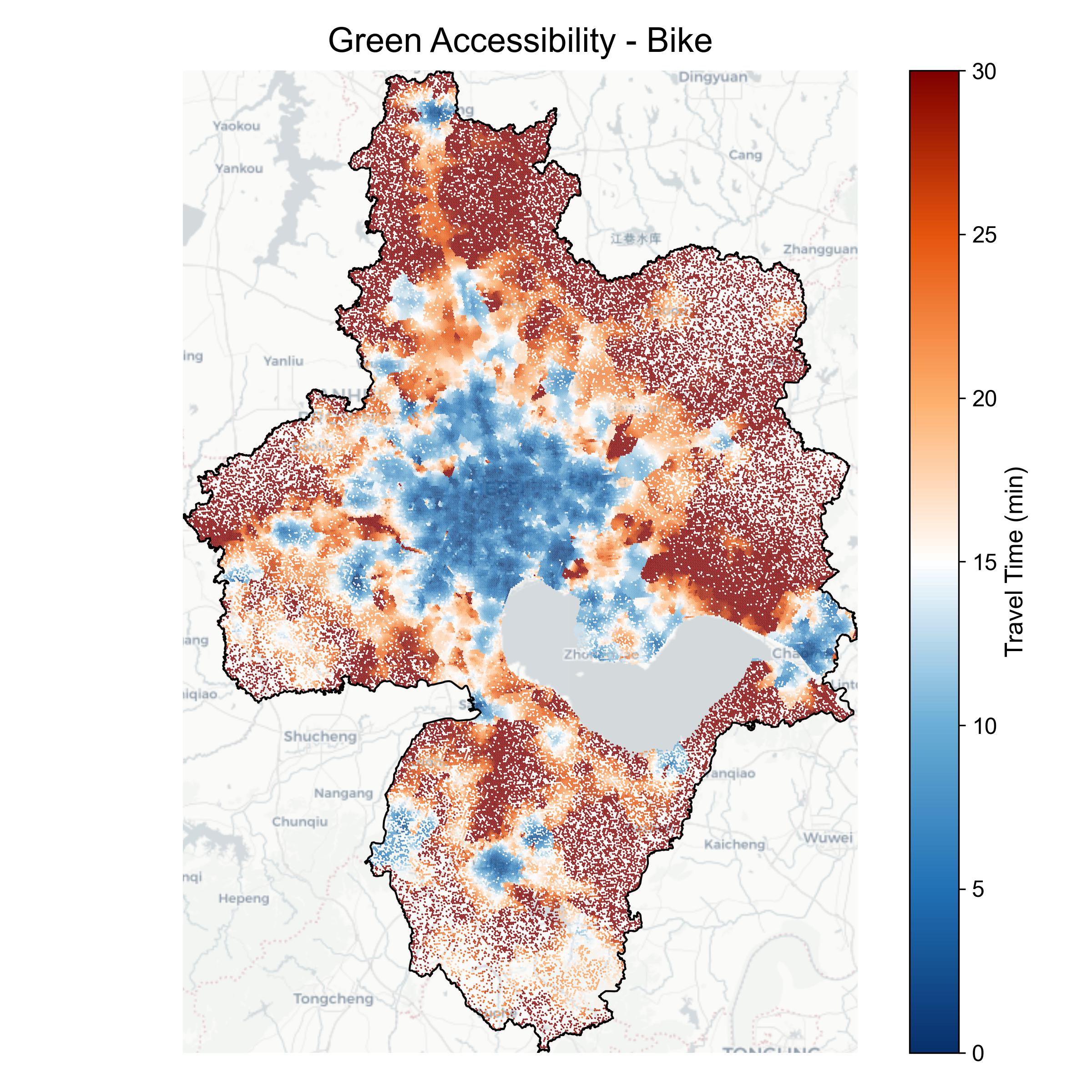}
\end{minipage}
\hspace{0.02\textwidth}
\begin{minipage}[t]{0.4\textwidth}
    \textbf{(b)}\\[2pt]
    \includegraphics[width=\linewidth]{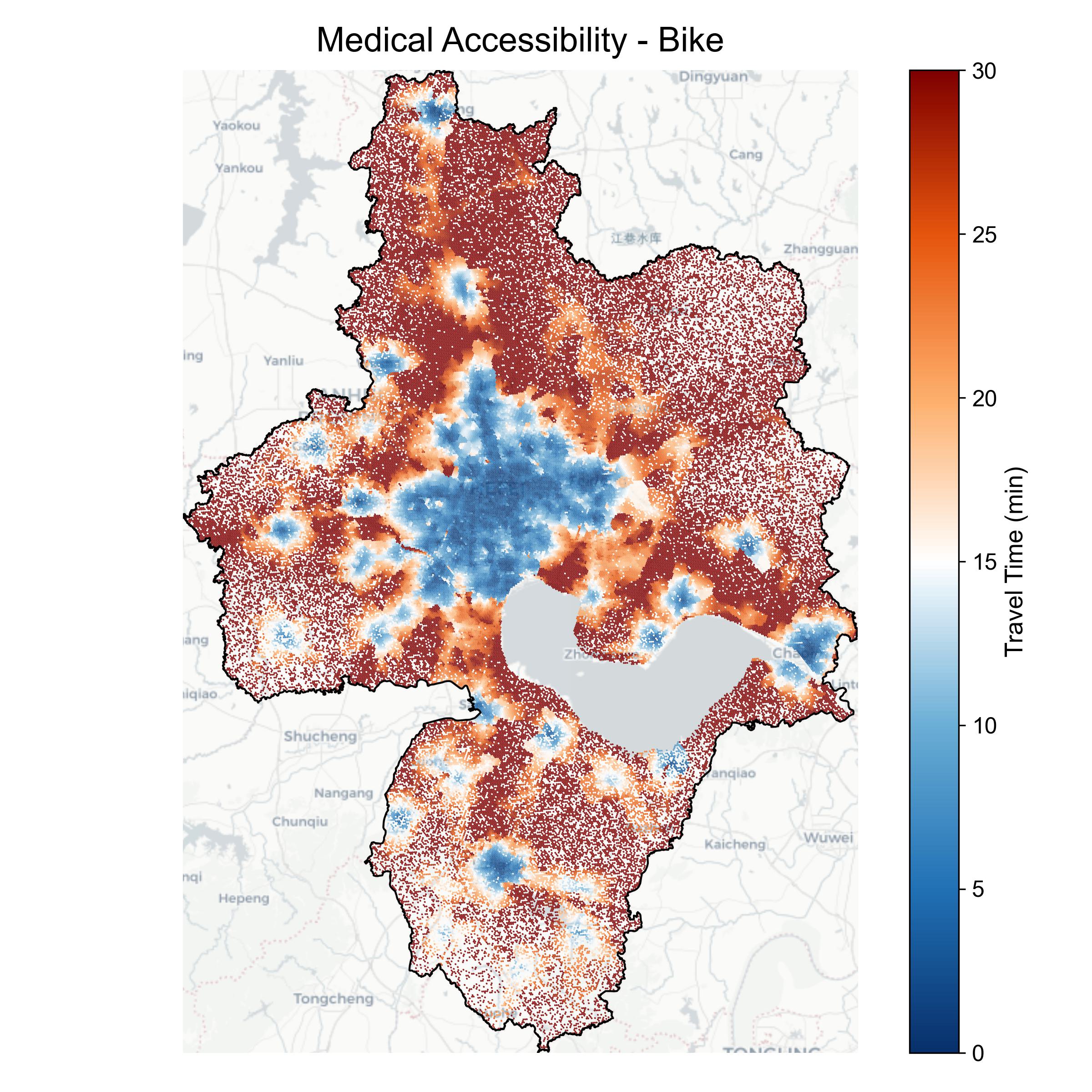}
\end{minipage}
\caption{Spatial distribution of proximity time.
(a) Green space accessibility by cycling.
(b) Medical service accessibility by cycling.}
\label{fig:accessibility_maps_bike}
\end{figure}

Fig.~\ref{fig:accessibility_maps}a,b maps proximity time for green spaces and medical services under walking. Under walking, short travel times are concentrated in the urban core, whereas peripheral and newly developed zones exhibit consistently higher travel times. Cycling (Fig.~\ref{fig:accessibility_maps_bike}) expands low- and moderate-travel-time areas outward from the core and increases spatial continuity of accessible zones. Nevertheless, several peripheral areas remain characterised by relatively long travel times even under cycling.

Fig.~\ref{fig:accessibility_maps}c shows population-weighted cumulative distribution functions (CDFs) of nearest-facility travel time. Across both facility types, cycling is associated with shorter travel times for a given cumulative population share, indicating a systematic improvement in accessibility. %The improvement is particularly pronounced in the upper tail, suggesting that cycling reduces long-distance disadvantage more strongly than short-distance differences.

% The walking medical CDFs lies slightly above green-space CDFs within short travel times (approximately the first 15--20 minutes), implying that a larger share of the population can reach medical facilities quickly. Under cycling, the green-space distributions become comparable to, and in parts of the distribution slightly outperform, those of medical services.

For the same facility type and mode, workplace-based CDFs are shifted to the right of residential-based CDFs, indicating longer nearest-facility travel times during daytime exposure. %The residential--workplace gap is more pronounced under walking and for green spaces. 
However, changes in the accessibility patterns do not necessarily translate into changes in the provision of services per person, as proximity time does not account for the intensity of competition among exposed populations; we are going to examine these results next.

\subsection{Per-capita provision within a 15-minute exposure window: the competition dimension}
\label{sec:facility_density}
\begin{figure}[H]
\centering
% Row 1: green spaces (widths proportional to aspect ratio for equal heights)
\begin{minipage}[t]{0.236\textwidth}
    \textbf{(a)}\\[2pt]
    \includegraphics[width=\linewidth]{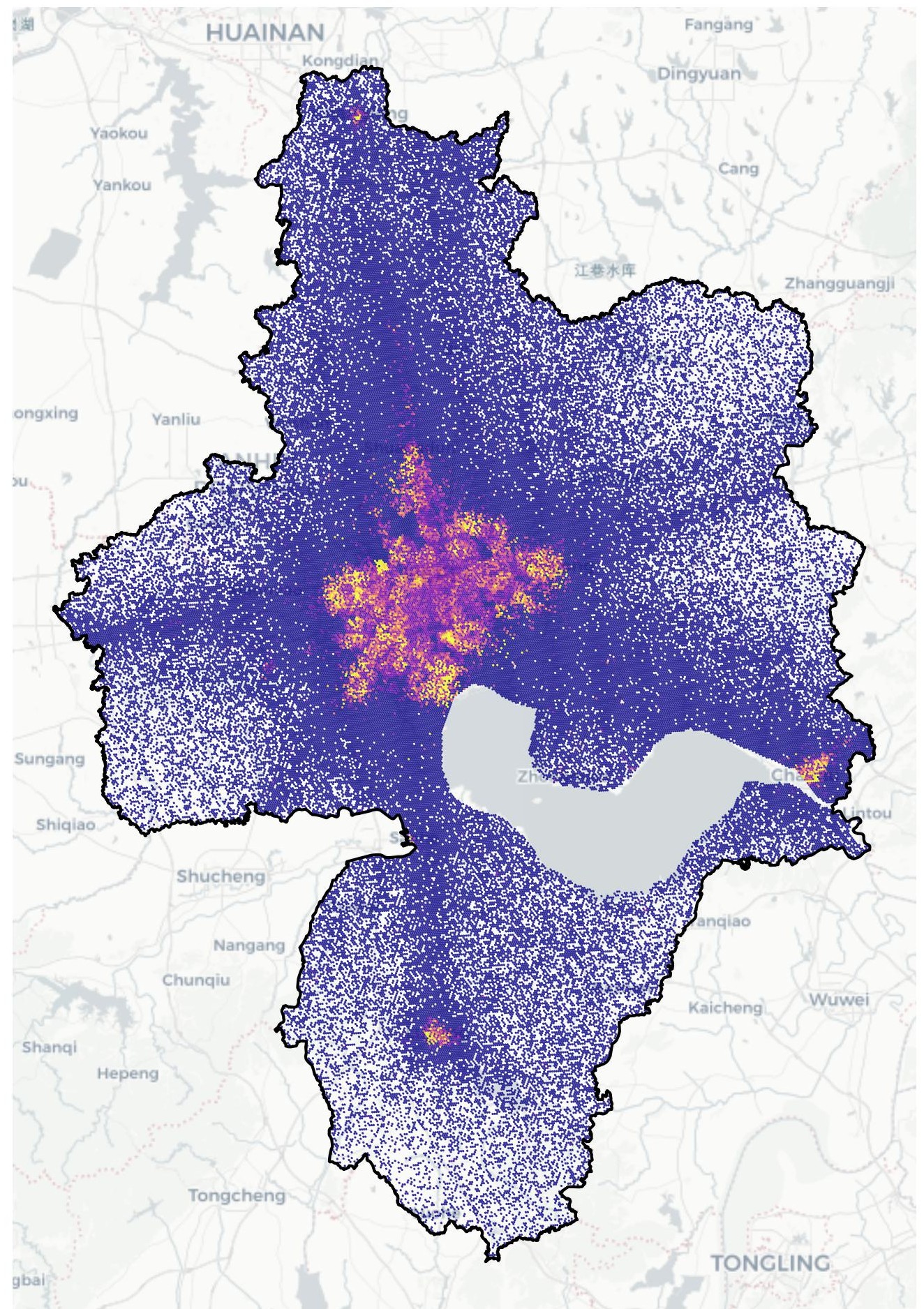}
\end{minipage}%
\hspace{0.015\textwidth}%
\begin{minipage}[t]{0.308\textwidth}
    \textbf{(b)}\\[2pt]
    \includegraphics[width=\linewidth]{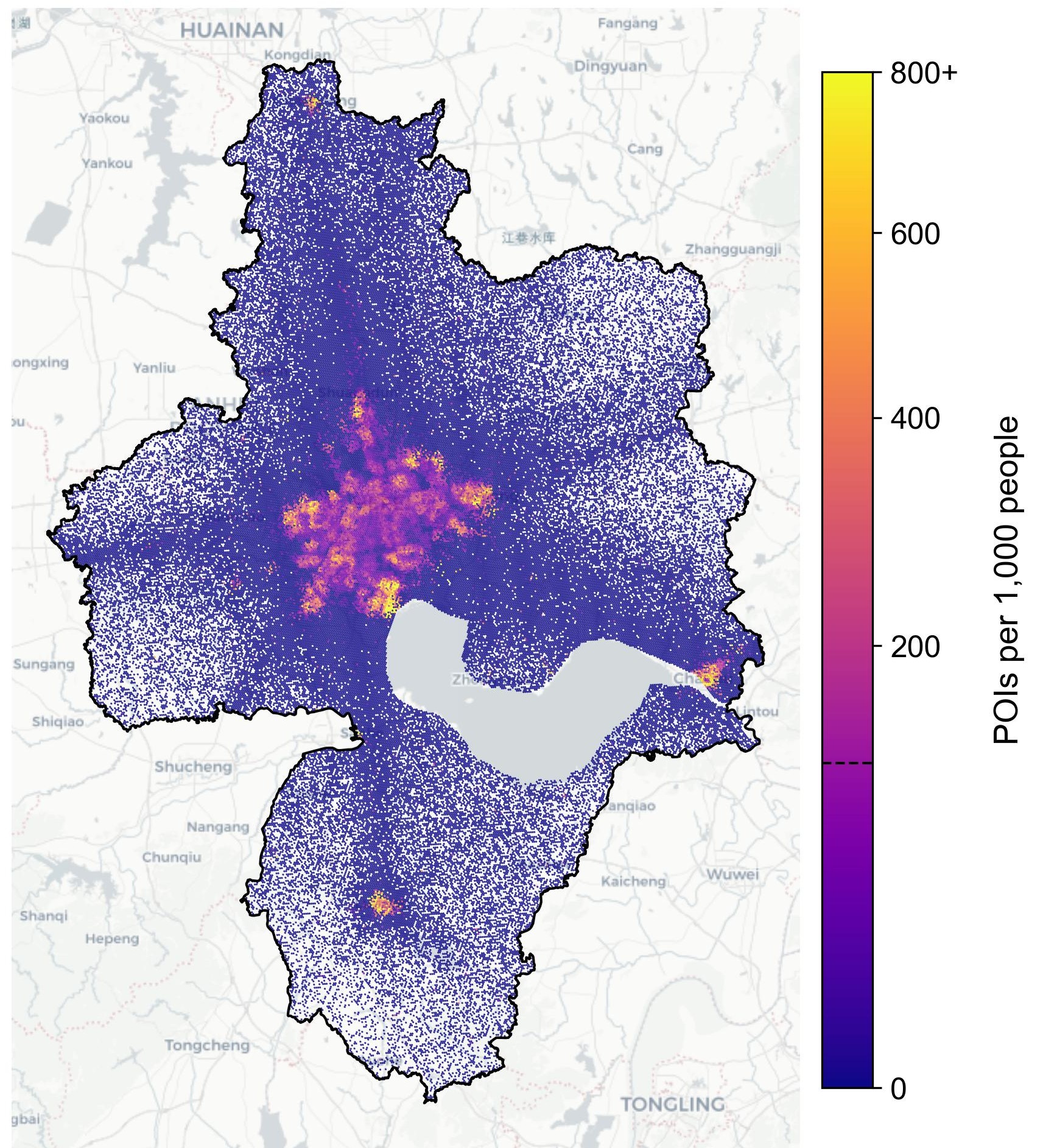}
\end{minipage}%
\hspace{0.015\textwidth}%
\begin{minipage}[t]{0.397\textwidth}
    \textbf{(c)}\\[2pt]
    \includegraphics[width=\linewidth]{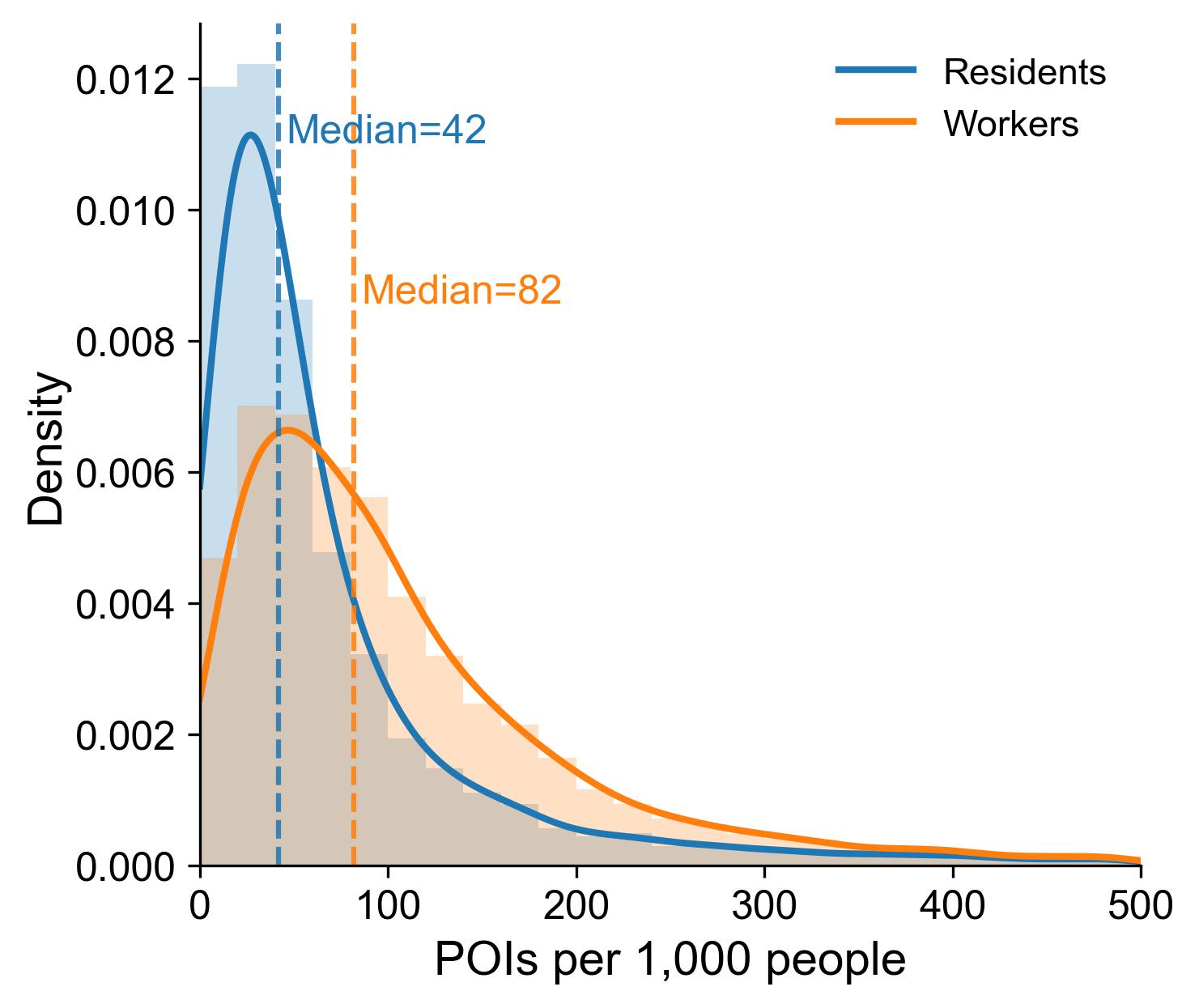}
\end{minipage}

\medskip

% Row 2: medical facilities
\begin{minipage}[t]{0.240\textwidth}
    \textbf{(d)}\\[2pt]
    \includegraphics[width=\linewidth]{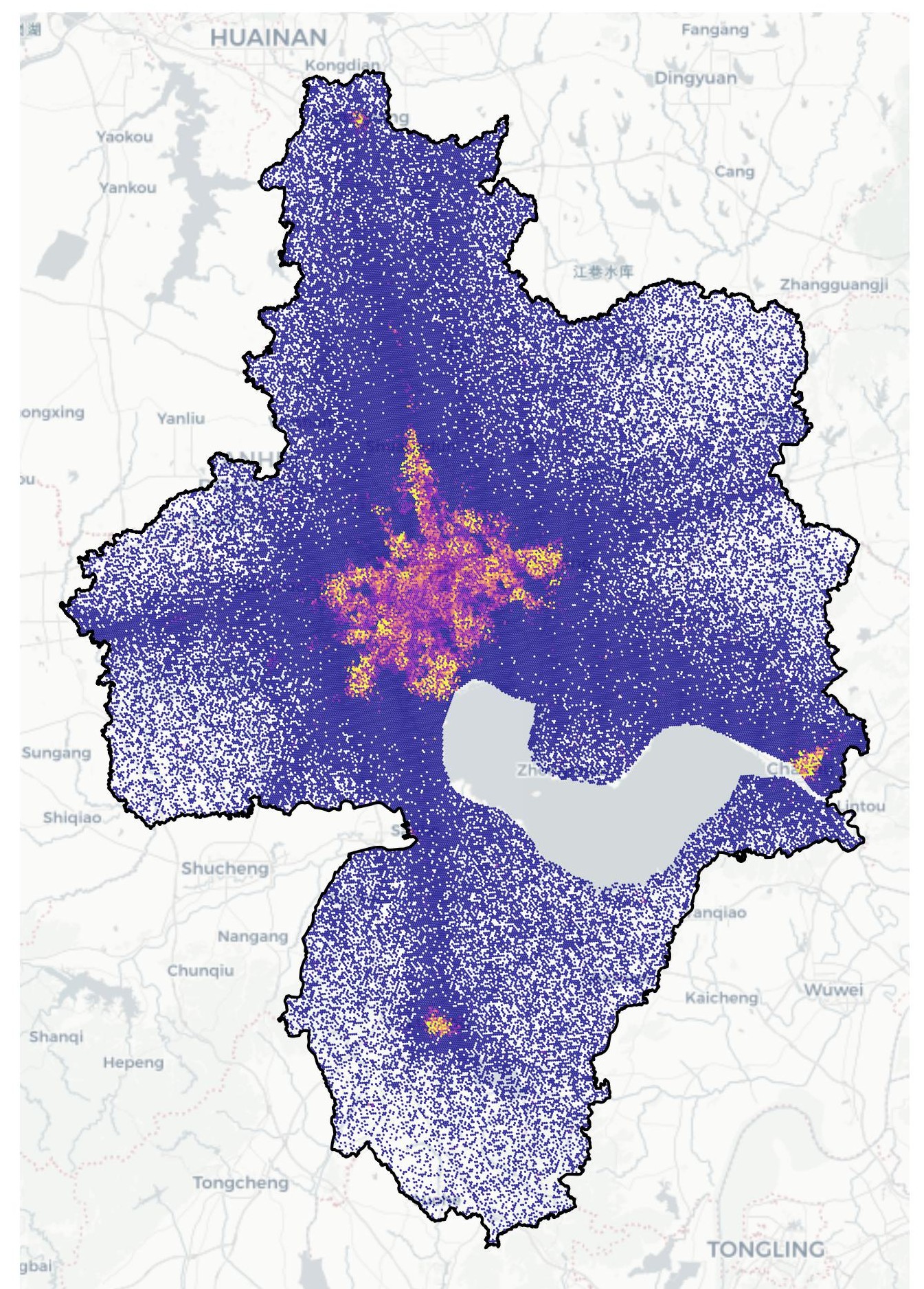}
\end{minipage}%
\hspace{0.015\textwidth}%
\begin{minipage}[t]{0.302\textwidth}
    \textbf{(e)}\\[2pt]
    \includegraphics[width=\linewidth]{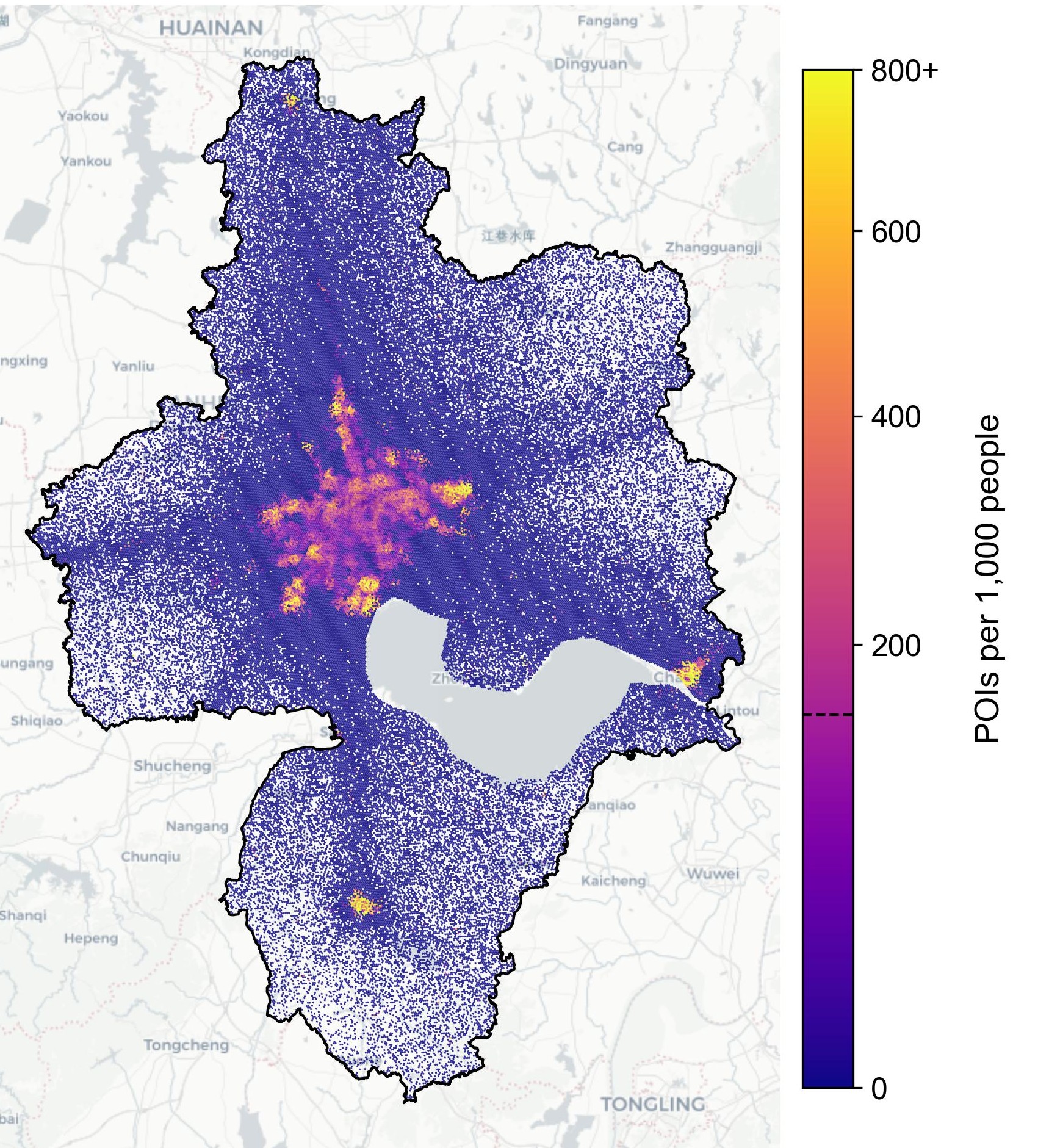}
\end{minipage}%
\hspace{0.015\textwidth}%
\begin{minipage}[t]{0.398\textwidth}
    \textbf{(f)}\\[2pt]
    \includegraphics[width=\linewidth]{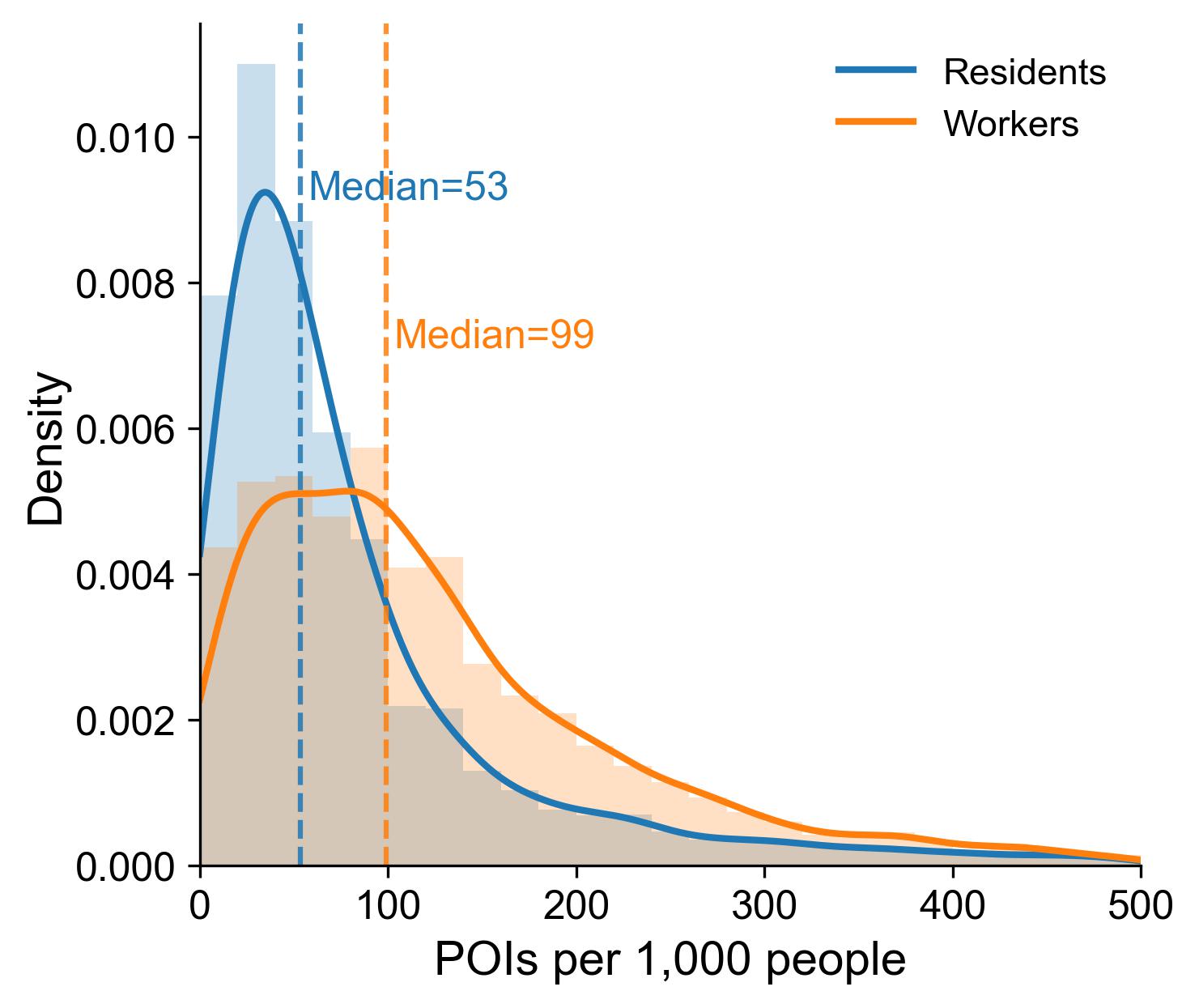}
\end{minipage}
\caption{\textbf{Per-capita provision within a 15-minute travel-time window} (facilities per 1{,}000 people), aggregated to H3 grids.
Upper row --- green spaces: (a) per 1{,}000 residents (home locations); (b) per 1{,}000 workers (workplace locations); (c) kernel density of per-capita green space provision across H3 cells, comparing residents (blue) and workers (orange), with population-weighted medians shown as dashed lines.
Lower row --- medical facilities: (d) per 1{,}000 residents (home locations); (e) per 1{,}000 workers (workplace locations); (f) kernel density of per-capita medical provision across H3 cells, comparing residents (blue) and workers (orange).
The gray area in maps (a), (b), (d), (e) corresponds to Chaohu Lake, the largest lake in Hefei.}
\label{fig:per_capita_provision}
\end{figure}

Fig.~\ref{fig:per_capita_provision} reports the number of facilities reachable within a 15-minute travel-time window per 1{,}000 people, computed separately for residential and workplace populations. %This indicator captures effective service availability conditional on both spatial reach and population concentration, thereby reflecting competition-adjusted provision rather than physical proximity alone.
Across both medical and green POIs, per-capita provision exhibits a pronounced core-periphery gradient. Higher values are concentrated in central districts, forming contiguous clusters of elevated provision, whereas peripheral and newly developed areas display persistently low values. These low-provision zones appear as spatially continuous belts rather than isolated pockets, indicating systematic disparities in effective service availability.

Interestingly, though, the highest values of per capita availability of services are not found in the centre, but in intermediate areas between core areas and the isolated peripheries. This result is clear in the case of resident population (Fig.~\ref{fig:per_capita_provision}a,d), and it is only partially mitigated when considering the working population (Fig.~\ref{fig:per_capita_provision}b,e); indeed, the abundance of workplaces in some areas drives a high provision of services considering only the resident population, but the provision stays still high when considering the flows of workers.

% Workplace-based provision surfaces (Fig.~\ref{fig:per_capita_provision}b,d) appear more spatially uneven than residential-based surfaces (Fig.~\ref{fig:per_capita_provision}a,c). In several employment-concentrated districts, provision levels remain only moderate despite dense facility presence, consistent with strong daytime population inflows that dilute per-capita availability within the 15-minute window.

The kernel density plots (panels c and f) depict the distribution of per capita provision of services per hexagon. The competition effect and the workers-residents differences are also clear from here. In both panels, the population-weighted median for residents lies below that for workers, indicating that residents are disproportionately located in areas with below-average per-capita provision relative to the working population. The divergence between these distributions reflects the tendency of the residential population to concentrate in districts where competition pressure reduces effective service availability. Service distribution is closer to the spatial distribution of the working population, in contrast with the proximity ideal in which residential neighbourhoods are mixed-use and adequately provided with amenities.

Taken together, these results suggest that considering the competition of actual population we can better understand the effective service availability within a 15-minute window compared to considering just facility density metrics. Central areas may combine dense facilities with intense competition, while peripheral areas may experience lower competition but limited facility reach. The spatial divergence between accessibility and competition-adjusted provision motivates a two-dimensional assessment of compounded disadvantage in the following section.

% These findings suggest that effective access is jointly shaped by spatial reach and population-induced dilution effects. Improvements in one dimension cannot compensate for structural deficiencies in the other.

\subsection{Double disadvantage: joint identification and mode-shift effects}
\label{sec:double_disadvantage}

\begin{figure}[H]
\centering
\begin{minipage}[t]{0.381\textwidth}
    \textbf{(a)}\\[2pt]
    \includegraphics[width=\linewidth]{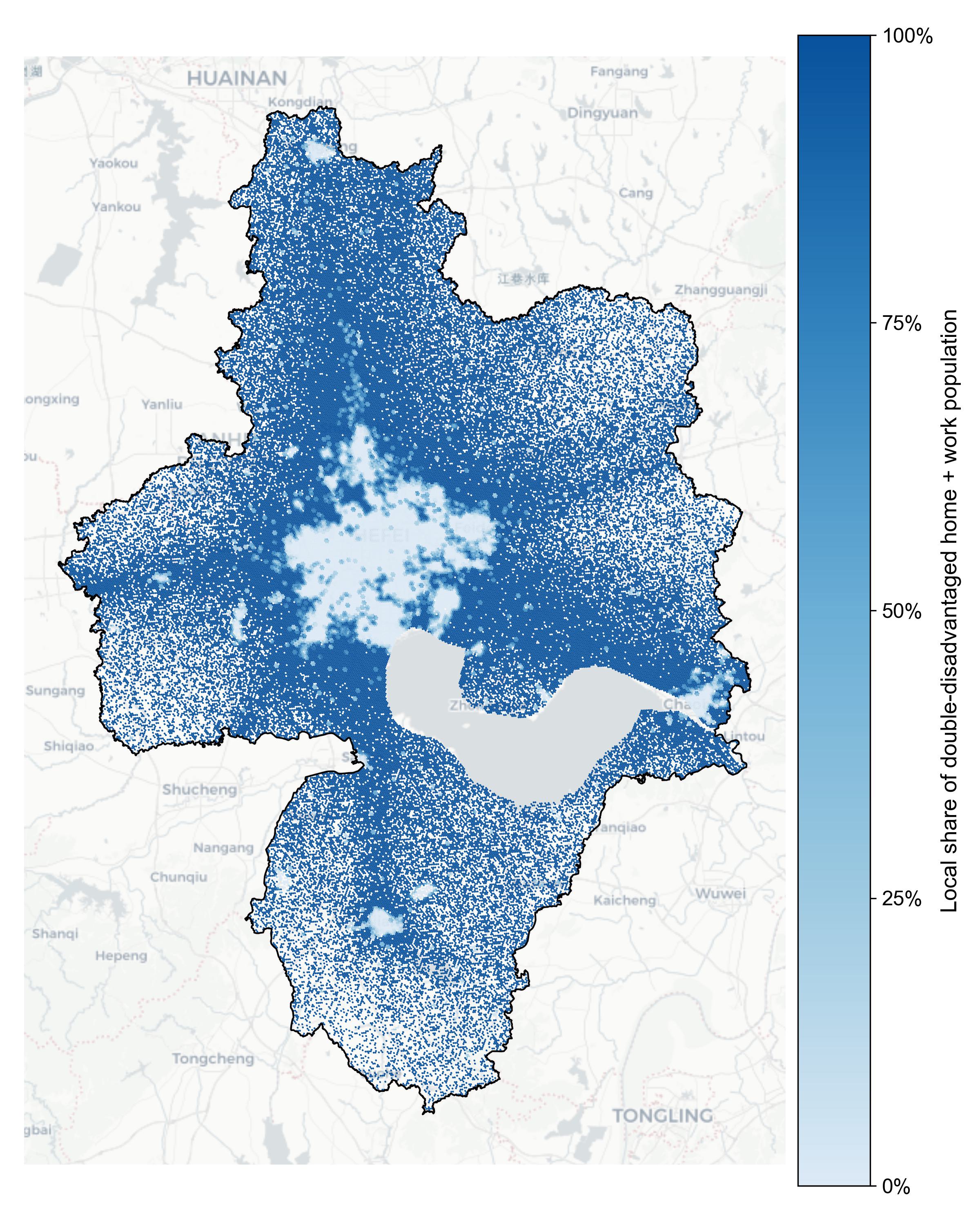}
\end{minipage}
\hspace{0.02\textwidth}
\begin{minipage}[t]{0.540\textwidth}
    \textbf{(b)}\\[2pt]
    \includegraphics[width=\linewidth]{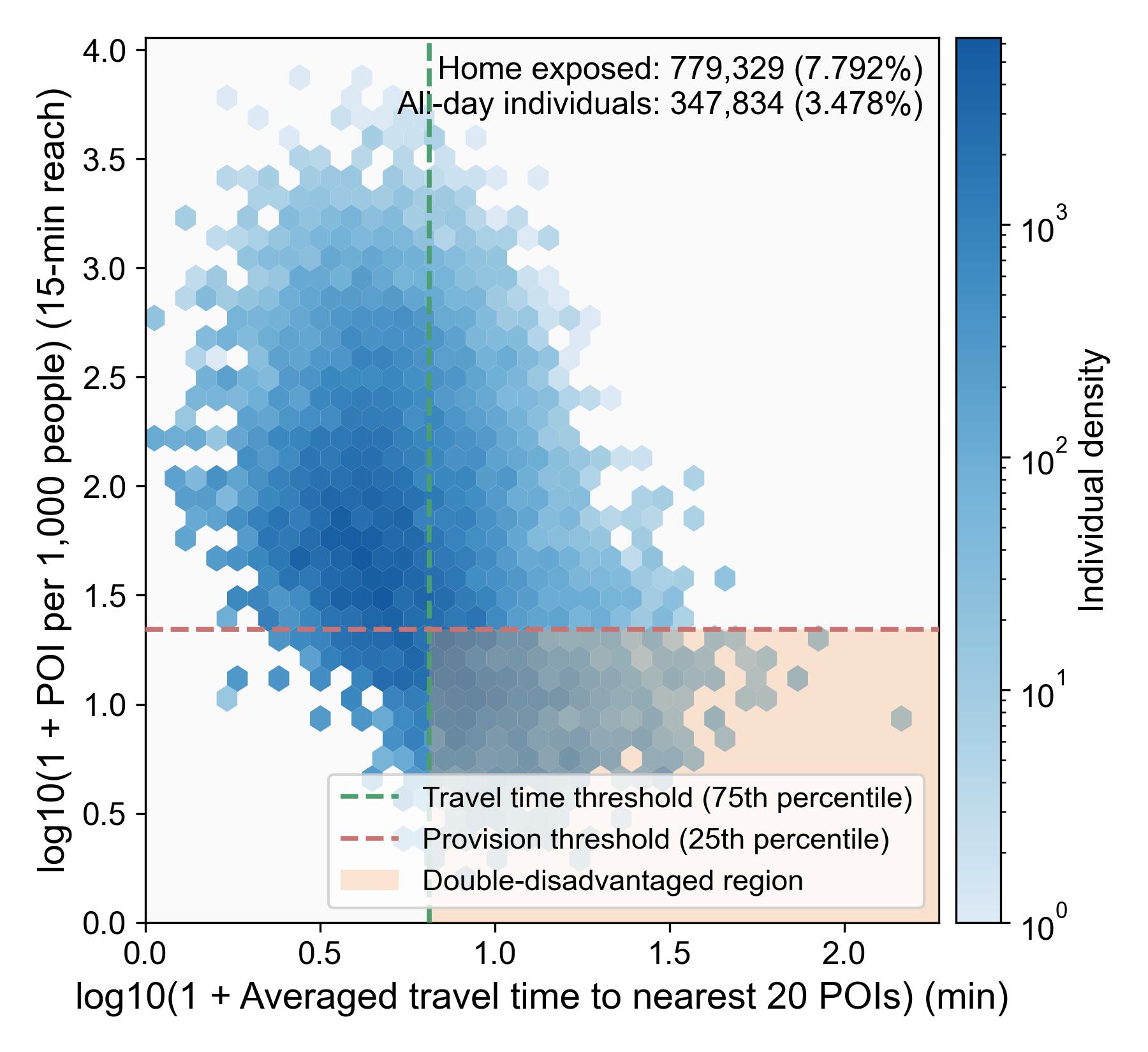}
\end{minipage}

\medskip

\begin{minipage}[t]{0.381\textwidth}
    \textbf{(c)}\\[2pt]
    \includegraphics[width=\linewidth]{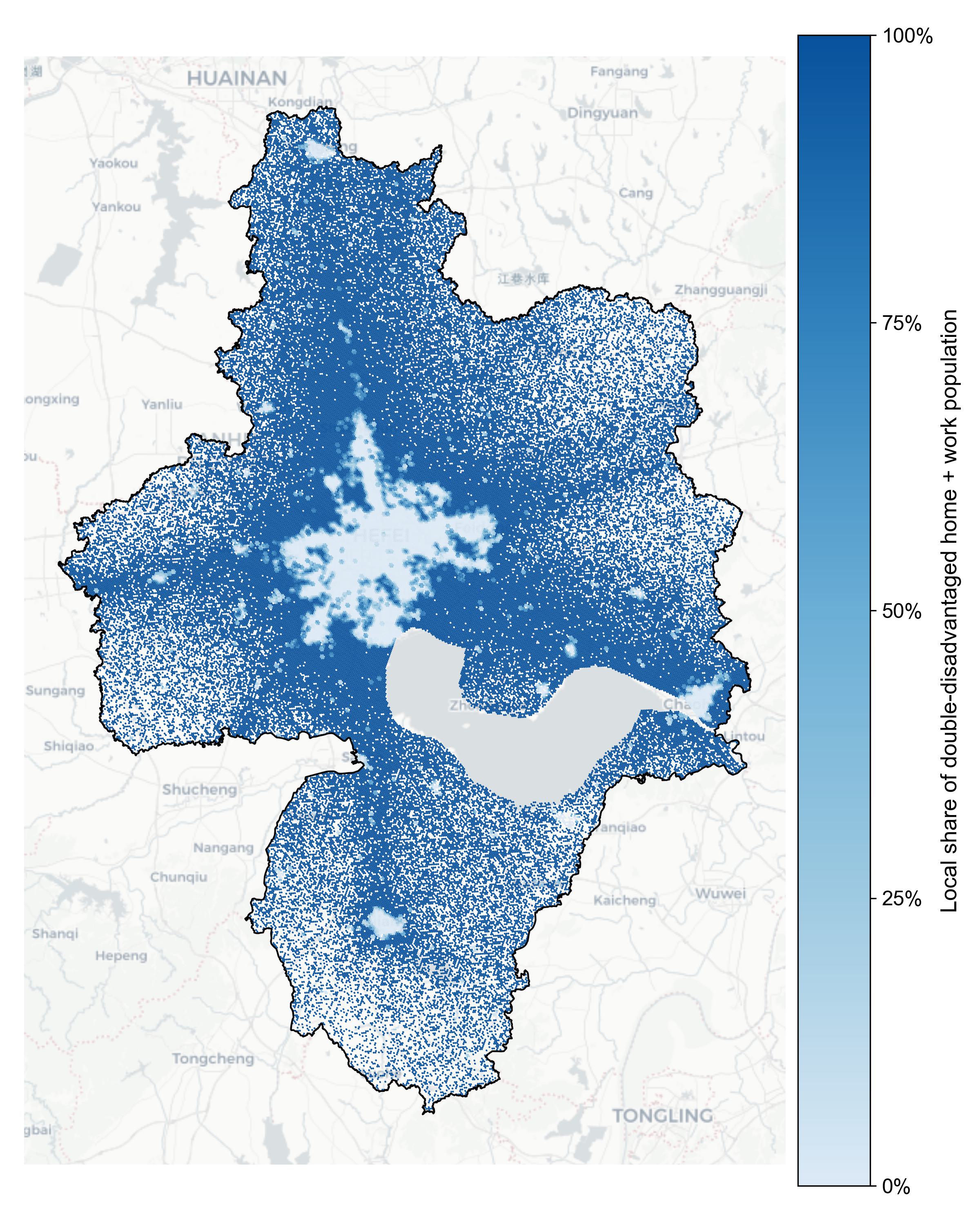}
\end{minipage}
\hspace{0.02\textwidth}
\begin{minipage}[t]{0.540\textwidth}
    \textbf{(d)}\\[2pt]
    \includegraphics[width=\linewidth]{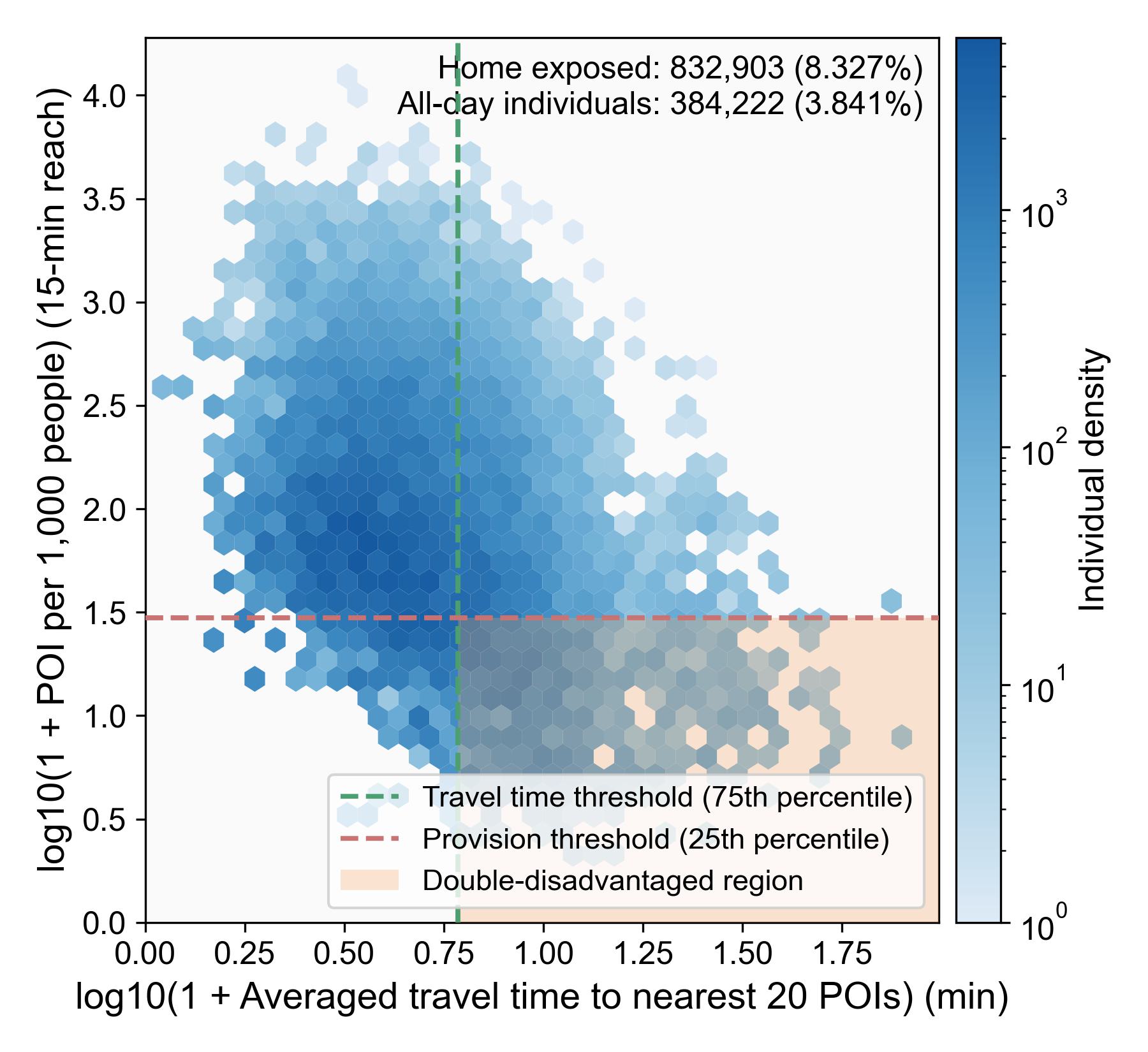}
\end{minipage}
\caption{\textbf{Double disadvantage: spatial distribution and individual-level identification under cycling accessibility conditions.}
Left column: spatial distribution of the all-day double-disadvantaged population share.
Right column: hexbin plot of proximity time versus per-capita provision in the residential context, coloured by number of individuals; dashed lines mark the quartile thresholds used to define double disadvantage.
(a,b) Green spaces --- statistics in (b) correspond to the ``Green -- Bike -- Home'' row in Table~1.
(c,d) Medical services --- statistics in (d) correspond to the ``Medical -- Bike -- Home'' row in Table~1.
The remaining mode--context combinations are presented in the Appendix (Fig.~\ref{fig:appendix_green_hexbin} for green spaces; Fig.~\ref{fig:appendix_medical_hexbin} for medical facilities).}
\label{fig:double_disadvantage}
\end{figure}

Double disadvantage is identified at the H3 grid level by jointly applying two distribution-based criteria within each population context: (i) nearest-facility travel time in the worst population-weighted quartile; and (ii) per-capita provision in the lowest population-weighted quartile (see Methods). Fig.~\ref{fig:double_disadvantage}b,d illustrate the individual-level distribution of travel time and per-capita provision under the same log-based thresholds.

The individual-level hexbin plots show a weak correlation between the two dimensions (Spearman = -0.344; Pearson = -0.123 for green spaces). Low travel times can coincide with low per-capita provision, and conversely, longer travel times do not necessarily imply low competition-adjusted availability. This weak alignment indicates that the two dimensions capture distinct aspects of service access. Fig.~\ref{fig:poi_home_vs_work} in the Appendix further illustrates the context-dependence of provision: at the H3 level, workplace-based green-space provision is generally higher than residential-based provision, yet the individual-level relationship between the two contexts remains weak, reflecting the spatial mismatch between where people live and where they work.

Fig.~\ref{fig:double_disadvantage}a,c map the spatial distribution of population exposed to all-day double disadvantage, where ``all-day'' refers to individuals located in H3 cells that are classified as disadvantaged in both residential and workplace contexts.

\begin{figure}[H]
\centering
\begin{minipage}[t]{0.29\textwidth}
    \textbf{(a1)}\\[2pt]
    \includegraphics[width=\linewidth]{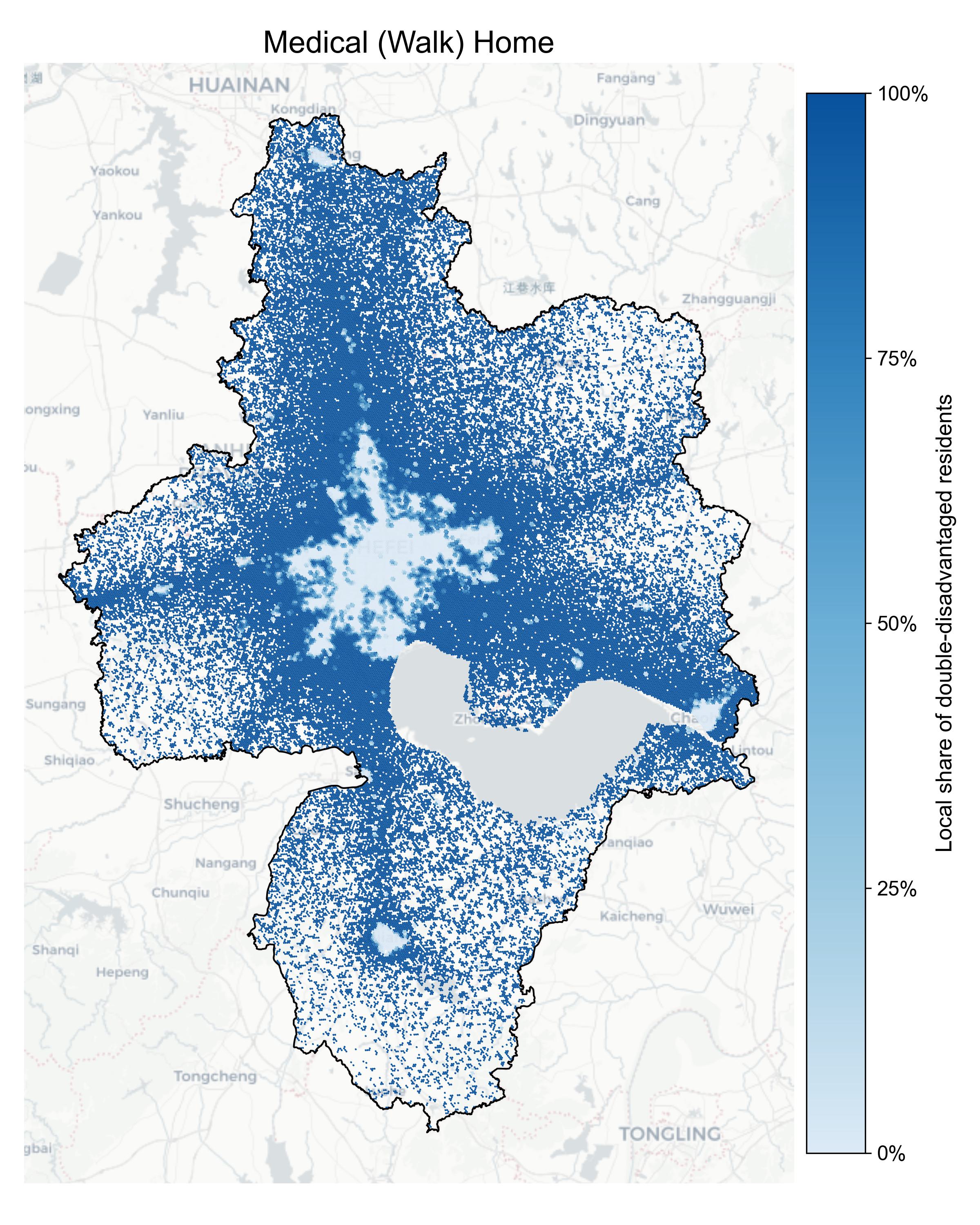}
\end{minipage}
\hspace{0.015\textwidth}
\begin{minipage}[t]{0.29\textwidth}
    \textbf{(b1)}\\[2pt]
    \includegraphics[width=\linewidth]{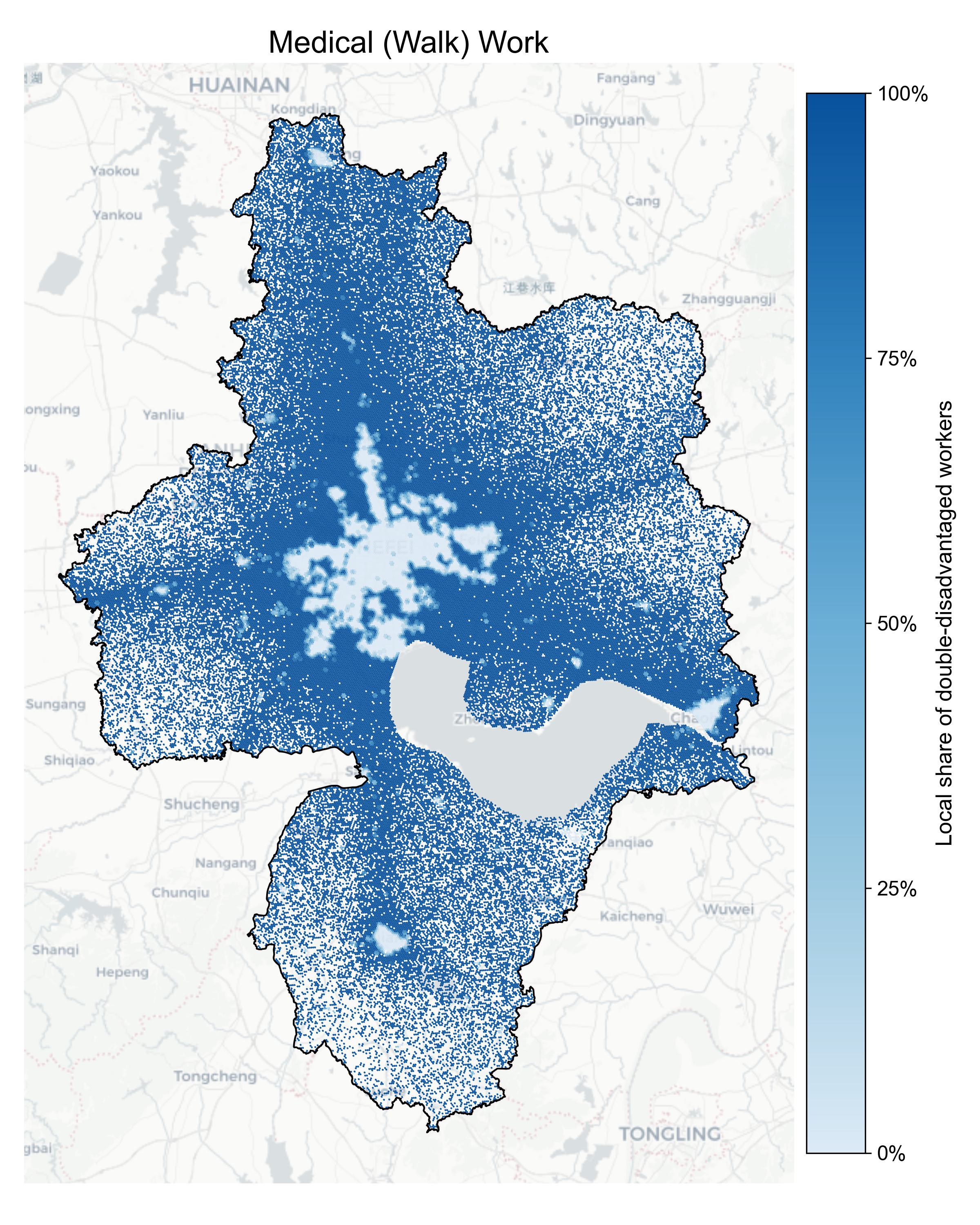}
\end{minipage}
\hspace{0.015\textwidth}
\begin{minipage}[t]{0.29\textwidth}
    \textbf{(c1)}\\[2pt]
    \includegraphics[width=\linewidth]{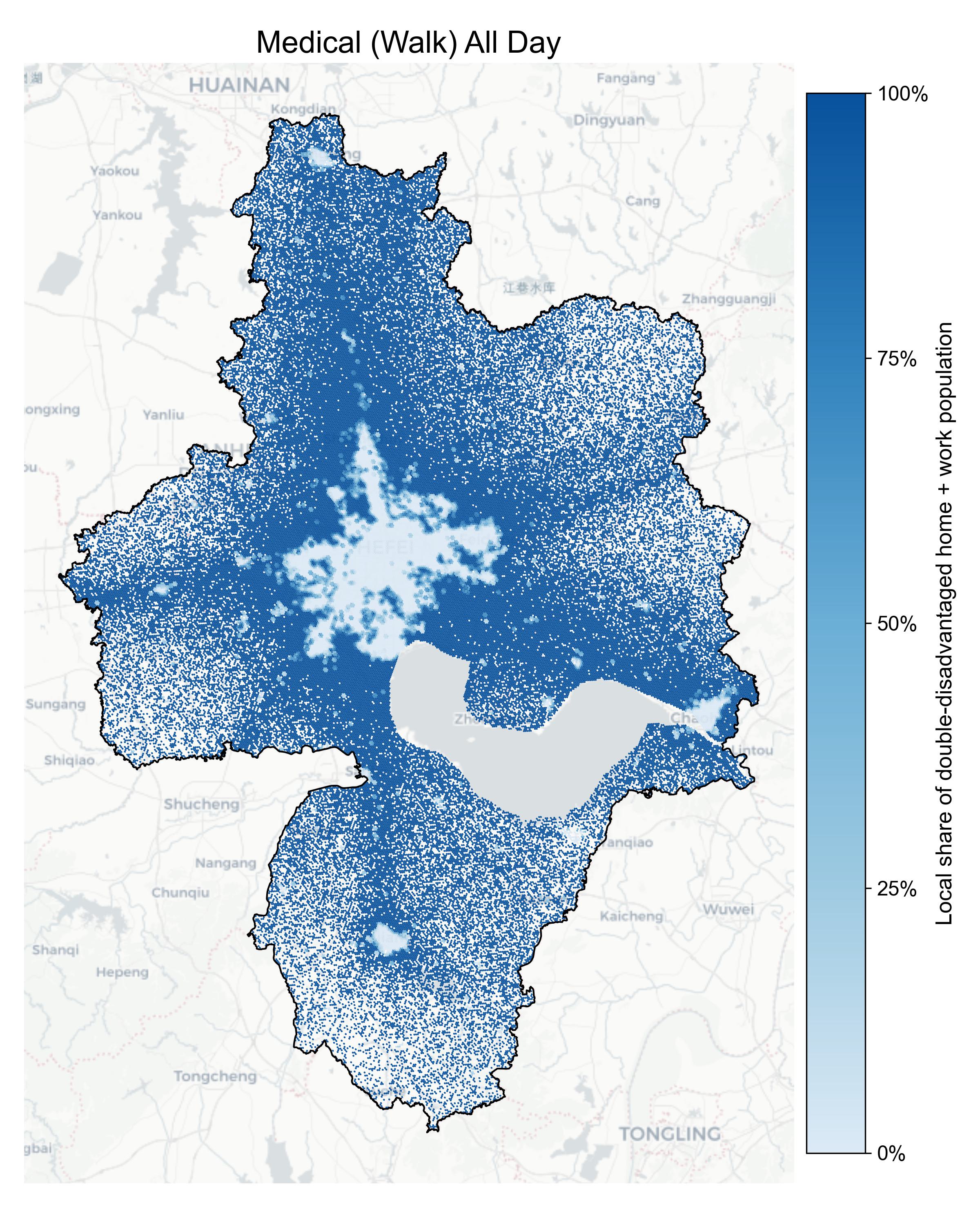}
\end{minipage}

\begin{minipage}[t]{0.29\textwidth}
    \textbf{(a2)}\\[2pt]
    \includegraphics[width=\linewidth]{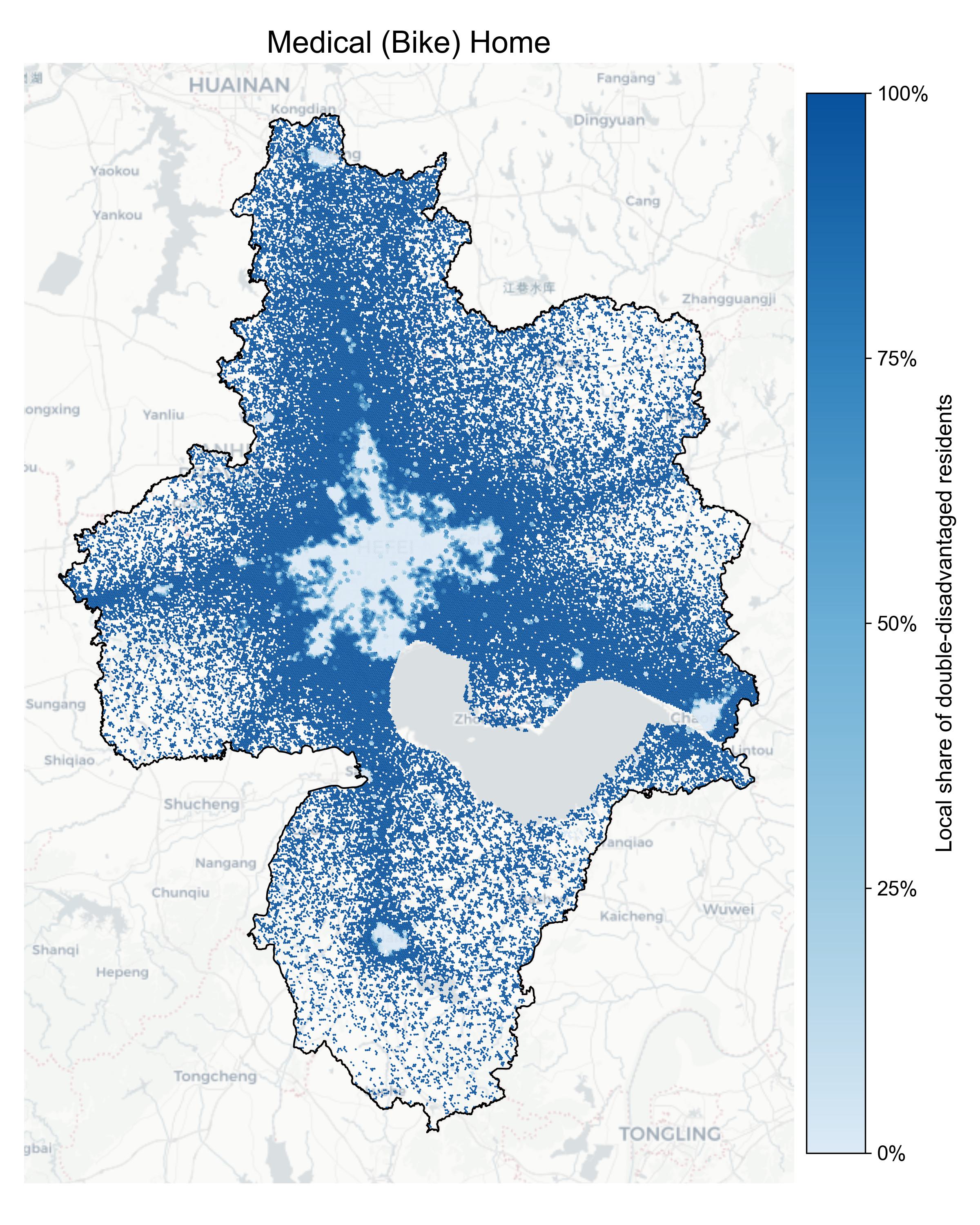}
\end{minipage}
\hspace{0.015\textwidth}
\begin{minipage}[t]{0.29\textwidth}
    \textbf{(b2)}\\[2pt]
    \includegraphics[width=\linewidth]{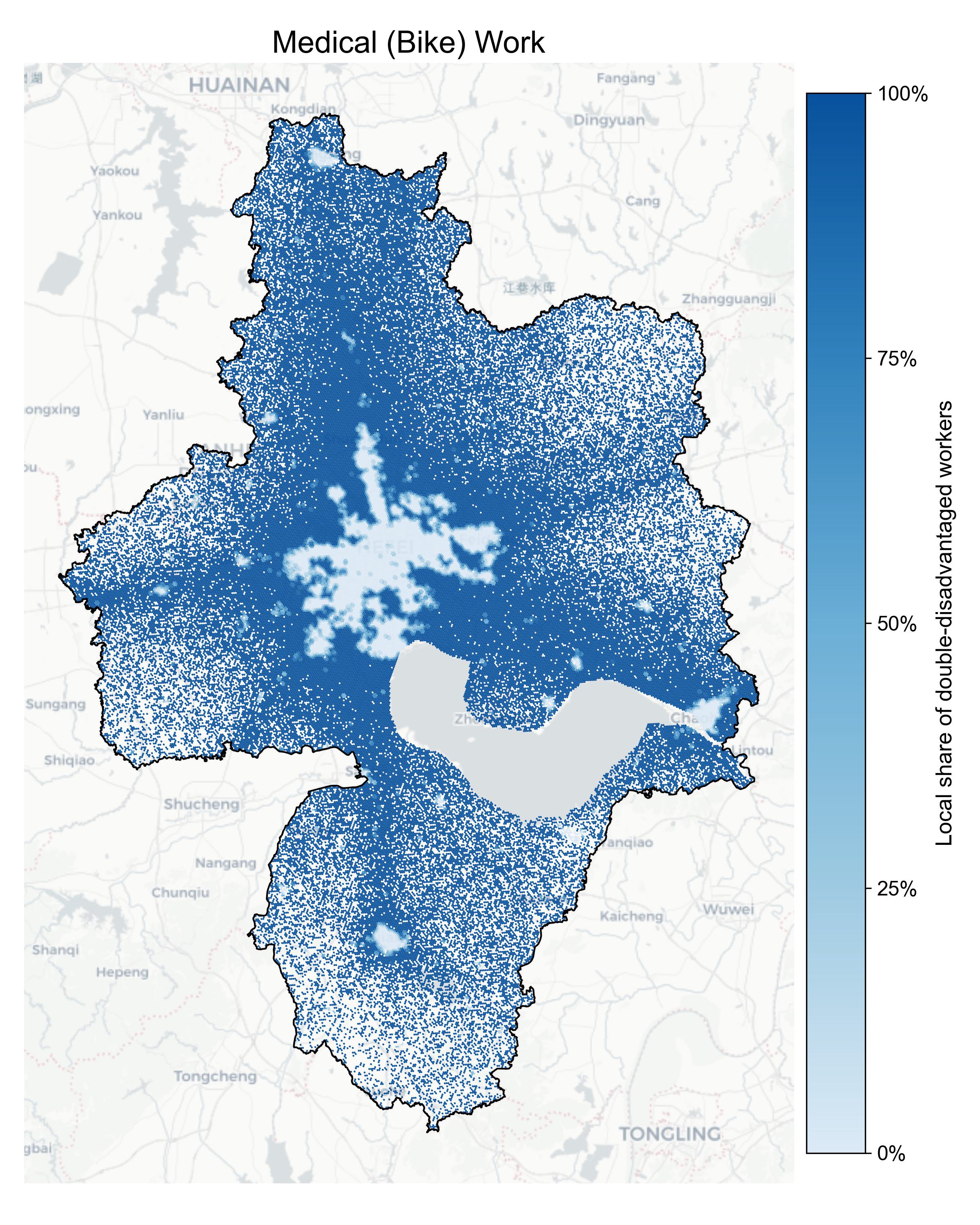}
\end{minipage}

\caption{Double-disadvantaged areas in access to medical services under walking and cycling conditions, expressed as the share of disadvantaged individuals relative to the corresponding reference population. (a) Residential exposure, normalised by residents. (b) Workplace exposure, normalised by workers. (c) All-day exposure, normalised by the combined residential–workplace reference population.}
\label{fig:appendix_medical_walk}
\end{figure}

Across both facility types, all-day double disadvantage is relatively limited in the central urban core, where residential and workplace locations generally exhibit good proximity to green spaces and medical services. Higher shares are concentrated around the inner suburban belt surrounding the core districts and extend outward along several radial corridors, rather than forming a simple center-periphery gradient. The complete spatial distributions across all exposure contexts and transport modes are provided in Appendix B (Fig.~\ref{fig:appendix_green_walk} for green spaces; Fig.~\ref{fig:appendix_medical_walk} for medical services).

\begin{table}
\centering
\caption{Extent of double disadvantage across facility types, modes, and exposure contexts. Percentages in parentheses are calculated relative to the corresponding residential or workplace population totals.}
\label{tab:double_disadvantage_extended}
\begin{tblr}{
  width = \linewidth,
  colspec = {Q[146]Q[110]Q[146]Q[150]Q[365]},
  cells = {c},
  hline{1-2,8,14} = {-}{},
}
Facility & Mode & Context & H3 cells & Population exposed\\
Medical & Walk & Home & 87,985 (92.061\%) & 840,812 (8.406\%)\\
Medical & Walk & Work & 88,541 (92.642\%) & 1,583,355 (15.830\%)\\
Medical & Walk & All-day & 87,521 (91.575\%) & 390,221 (3.901\%)\\
Medical & Bike & Home & 87,754 (91.819\%) & 832,903 (8.327\%)\\
Medical & Bike & Work & 88,333 (92.425\%) & 1,568,567 (15.683\%)\\
Medical & Bike & All-day & 87,347 (91.393\%) & 384,222 (3.841\%)\\
Green & Walk & Home & 86,123 (90.112\%) & 788,065 (7.879\%)\\
Green & Walk & Work & 86,894 (90.919\%) & 1,498,468 (14.982\%)\\
Green & Walk & All-day & 85,664 (89.632\%) & 353,375 (3.533\%)\\
Green & Bike & Home & 86,028 (90.013\%) & 779,329 (7.792\%)\\
Green & Bike & Work & 86,743 (90.761\%) & 1,486,582 (14.863\%)\\
Green & Bike & All-day & 85,544 (89.506\%) & 347,834 (3.478\%)
\end{tblr}
\end{table}

Table~\ref{tab:double_disadvantage_extended} quantifies the incidence of double disadvantage and the corresponding exposed population across facility types, transport modes, and activity contexts.

Across all scenarios, double disadvantage affects a large share of spatial units, with the number of disadvantaged H3 cells ranging between approximately 85{,}500 and 88{,}500. Despite this broad spatial footprint, population exposure shares are considerably smaller. Under medical services with walking, 840{,}812 individuals (8.41\%) are exposed in the residential context, and 1{,}583{,}355 individuals (15.83\%) are exposed in the workplace context. When both contexts are considered together, the number of all-day-exposed individuals declines to 390{,}221 (3.90\% of the analytical sample).

A similar pattern is observed for green spaces. Residential exposure under walking reaches 788{,}065 individuals (7.88\%), while workplace exposure reaches 1{,}498{,}468 individuals (14.98\%). However, the number of individuals simultaneously exposed in both contexts decreases to 353{,}375 (3.53\%). These numbers highlight that although disadvantaged grid cells are widespread, the subset of individuals experiencing compounded disadvantage across daily activity contexts is considerably smaller.

Comparing transport modes, differences between walking and cycling remain modest. Even though the disadvantage is computed as a quartile, cycling slightly reduces both the number of disadvantaged H3 cells and the corresponding exposed population. 
Overall, the results indicate that double disadvantage affects a small but significant proportion of the population when both residential and workplace contexts are considered simultaneously, with more than 300{,}000 people experiencing double disadvantage across their entire commuting patterns.

\section{Discussion}

This study shows that inequities in access to health-supportive urban services are not fully captured by travel time alone. Instead, we propose a more nuanced description capturing the combined influence of mobility conditions, service availability, and differences in population exposure over time. Considering these dimensions together helps clarify some of the factors associated with persistent disadvantage in competition-constrained urban systems.

\subsection{Service inequity as a joint function of spatial separation and demand competition}

The weak and heterogeneous association between nearest-facility travel time and per-capita provision indicates that inequity in service access is fundamentally multidimensional. Spatial separation from services and demand-induced competition represent two distinct but interacting dimensions of disadvantage.

Peripheral areas often experience longer travel times due to spatial isolation, but may face relatively moderate competition. In contrast, employment-intensive central districts frequently combine short travel times with intense competition resulting from daytime population inflows. Double disadvantage arises where these two dimensions intersect: where spatial separation coincides with high levels of demand pressure or scarcity of services, even in relation to low population densities.

This two-dimensional structure has important implications for accessibility evaluation. Conventional single-indicator accessibility measures risk misclassifying densely populated areas with severe competition as well-served, while simultaneously overstating deprivation in areas with moderate demand but longer travel distances. By incorporating competition-adjusted service provision, the proposed framework reframes service inequity as a capacity-constrained accessibility problem rather than a purely distance-based one.

% In urban systems characterized by spatially concentrated demand, congestion-like externalities extend beyond transport networks to service provision systems. Under such conditions, improvements in mobility alone do not proportionally translate into improvements in effective access. Equity assessments must therefore explicitly incorporate demand competition to avoid overstating the distributive benefits of mobility-oriented policies.

When many people rely on the same services, access depends not only on how quickly they can reach them, but also on how much demand those services face. In these areas, shorter travel times may have limited benefits if services are already under pressure. Equity assessments should therefore consider both mobility and competition for services.

\subsection{Temporal exposure and hierarchical constraints}

A further finding concerns how population exposure changes over the day. Residential and workplace populations are distributed differently, creating different access conditions at night and during working hours. Employment centres may be well connected to services, but the concentration of workers during the day can increase demand and reduce effective availability.

Overall, the results show that the redistribution of population from residential areas to workplaces changes the geography of accessibility demand, but it does not remove disadvantage. In many cases, areas with poor residential access also face limited effective access in workplace contexts, especially where daytime population concentration increases competition for services. This suggests that workplace accessibility may reproduce, rather than compensate for, existing patterns of disadvantage.

These findings indicate that accessibility should be assessed across daily activity contexts rather than solely from residential locations. While home and work environments generate different demand patterns, disadvantaged populations often remain exposed to poor service access across both contexts. For planning, this means that mobility improvements and service provision strategies should account for how population shifts during the day reshape demand without necessarily reducing underlying inequalities.

\subsection{Policy implications}

Our findings suggest that persistent service disadvantage arises from the interaction between mobility conditions, facility distribution, and dynamic patterns of population exposure. While improvements in active mobility can mitigate travel-time barriers, they cannot by themselves reconfigure capacity allocation or reduce competition intensity in high-demand environments.

Several policy implications follow. First, equity evaluations should prioritise population-weighted outcomes rather than spatial coverage alone, since reductions in disadvantaged locations may not correspond to meaningful reductions in affected populations. Second, investments in active mobility should be complemented by adjustments in the spatial distribution and capacity of essential services, particularly in rapidly expanding urban fringe areas. Third, changes in population distribution between residential and workplace contexts should be explicitly considered, as daytime shifts in demand are non-negligible and can substantially alter patterns of effective access.

More generally, the results suggest that transport equity cannot be addressed solely through mobility-oriented interventions. In competition-constrained urban systems, distributive outcomes depend on coordinated planning across transport infrastructure, land-use development, and service provision.

\section{Conclusions}

This study establishes a population-based, temporally differentiated framework for assessing equity in access to health-supportive urban services within the 15-minute city paradigm. By integrating behaviour-derived population exposure, network-based travel times, and competition-adjusted per-capita provision, this approach moves beyond conventional, static residence-based models. Ultimately, it captures the critical structural interaction between human mobility, localised service capacity, and dynamic population distributions.

Our findings reveal a distinct decoupling between traditional time-based proximity and actual competition-adjusted service provision. While spatial distance dictates the physical ease of reaching a destination, the actual capacity of a facility, measured against real-time local demand densities, is what determines true access. Because local population densities oscillate dramatically throughout the day, service inequity is inherently multidimensional and temporally structured. Spatial separation and localised demand competition jointly shape effective access; during working hours, the massive influx of people into job centres drastically intensifies pressure on local services, shifting the burden of disadvantage away from purely residential zones. These dynamics underscore the need to embed both competition effects and time-dependent population exposure into future equity assessments.

From a policy perspective, these results prove that infrastructure and mobility improvements alone cannot bridge structural urban inequalities. Achieving true equity requires coordinated interventions that simultaneously optimise transport networks, scale up facility capacities, and realign service allocations with dynamic patterns of population exposure. The double-disadvantage framework proposed here offers a scalable, transferable methodology for identifying these shifting pockets of structural inequity and evaluating the distributive impacts of transport and land-use policies in rapidly urbanising environments.

Several limitations outline fruitful paths for future research. First, while large-scale mobile phone GPS data offer unprecedented spatial coverage, our sample is restricted to users with trackable home-to-work routines, potentially underrepresenting vulnerable populations with less stable daily trajectories. Second, service provision was modelled using facility counts without fully accounting for variations in institutional capacity or qualitative differences, such as the operational gap between a major regional hospital and a local general practitioner. Finally, while this analysis successfully isolates residential and workplace exposure, it does not fully capture fine-grained intra-day variations, weekend vs weekday dynamics, or non-work activity chains. Future extensions should incorporate richer behavioural trajectories, granular quality-of-service metrics, and group-specific demand profiles to more deeply map the moving target of urban equity.

\section{Acknowledgements}
% 72222021 田老师优青
% 72431006 马老师重点
% 72288101 高老师基础科学中心
% W2411064 姜老师国际合作
% BX20240033 郑老师博新
% 72401022 郑老师青年

This work is supported by the National Natural Science Foundation of China
(Grant Nos. 72222021, 72431006, and W2411064) and the China Scholarship Council
(Grant No. 202506250110).
\clearpage

\section*{Appendix A. Population data validation and activity patterns}

\section*{Appendix B. Double disadvantage maps}

\begin{figure}[H]
\centering
\textbf{\normalsize Green Spaces}\\[4pt]
\begin{minipage}[t]{0.45\textwidth}
    \textbf{(a)}\\[2pt]
    \includegraphics[width=\linewidth]{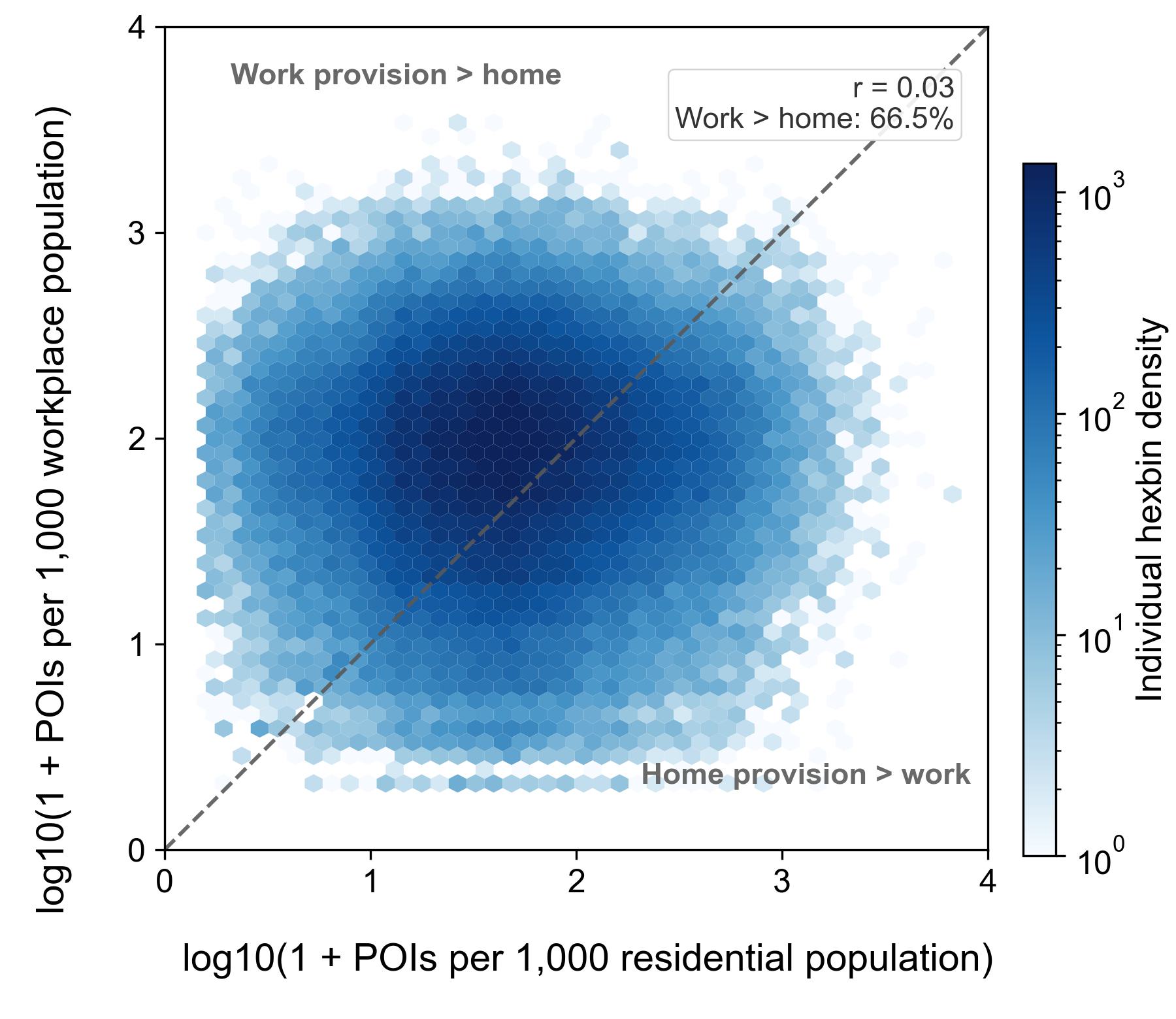}
\end{minipage}
\hspace{0.02\textwidth}
\begin{minipage}[t]{0.45\textwidth}
    \textbf{(b)}\\[2pt]
    \includegraphics[width=\linewidth]{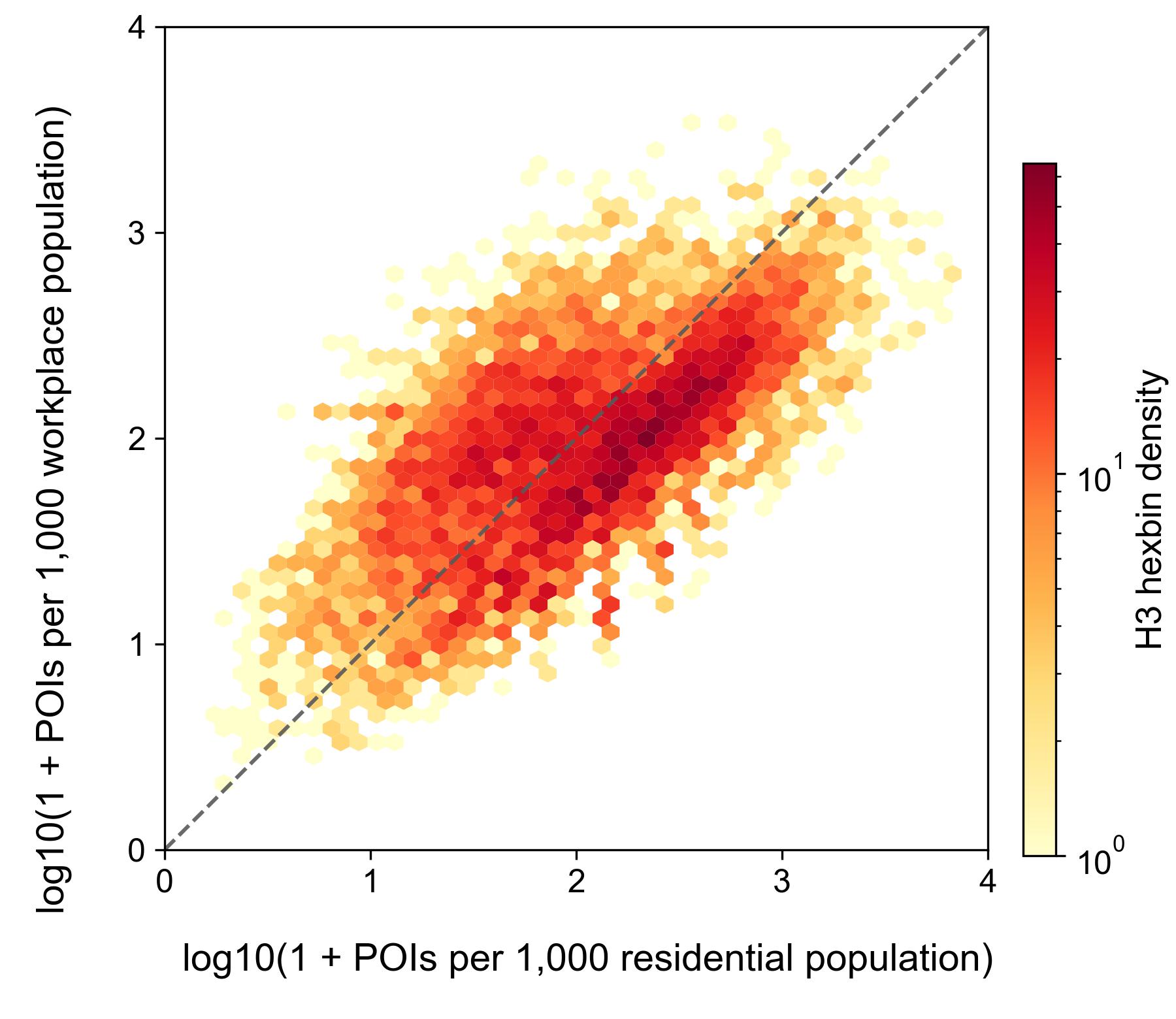}
\end{minipage}

\medskip

\textbf{\normalsize Medical POIs}\\[4pt]
\begin{minipage}[t]{0.45\textwidth}
    \textbf{(c)}\\[2pt]
    \includegraphics[width=\linewidth]{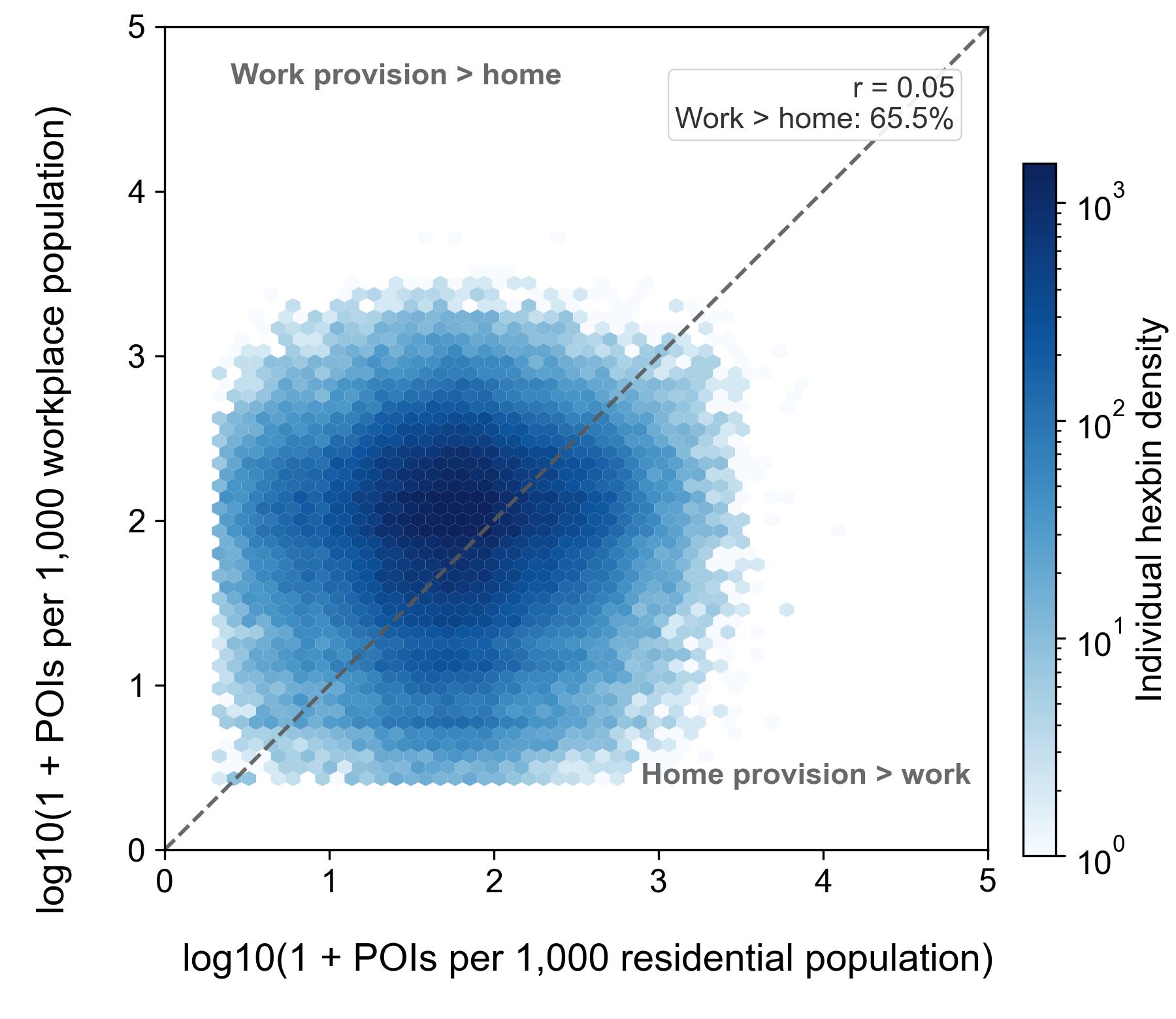}
\end{minipage}
\hspace{0.02\textwidth}
\begin{minipage}[t]{0.45\textwidth}
    \textbf{(d)}\\[2pt]
    \includegraphics[width=\linewidth]{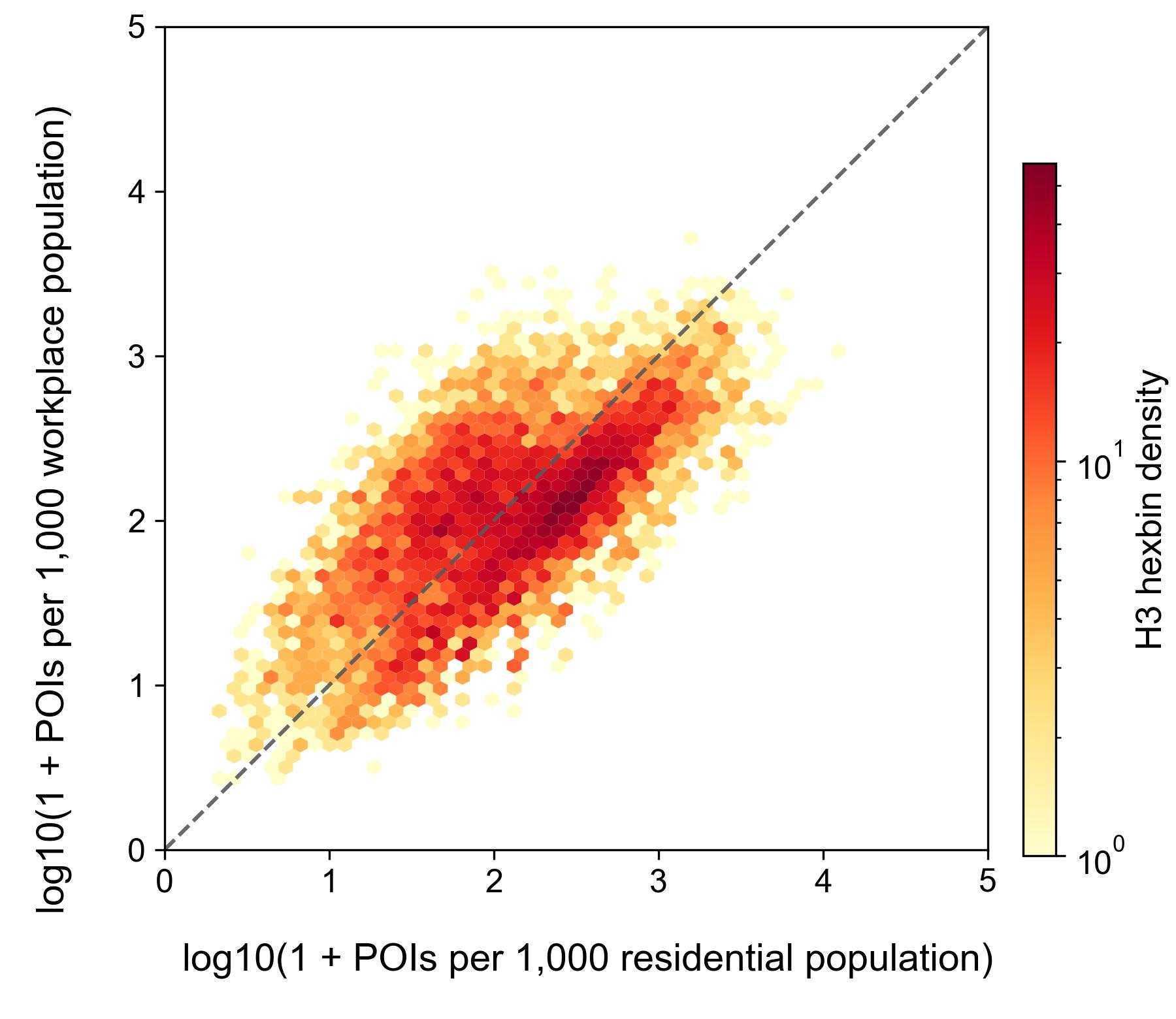}
\end{minipage}
\caption{
Hexbin plots comparing per-capita provision between residential and workplace contexts for green spaces (upper row, a--b) and medical facilities (lower row, c--d).
(a,c) Individual-level hexbin plot of residential- versus workplace-context provision; only individuals with positive provision in both contexts are included; each bin is coloured by the number of individuals.
(b,d) H3-cell-level hexbin plot of residential- versus workplace-context provision across grid cells.
In panels (a,b), provision is measured as green-space POIs reachable within 15 minutes per 1{,}000 population; in panels (c,d), the same metric is applied to medical facilities.
The dashed line denotes the 1:1 reference.
Workplace provision is generally higher than residential provision, whereas the individual-level relationship between the two contexts remains weak.
}
\label{fig:poi_home_vs_work}
\end{figure}

\begin{figure}[H]
\centering
\begin{minipage}[t]{0.45\textwidth}
    \textbf{(a)}\\[2pt]
    \includegraphics[width=\linewidth]{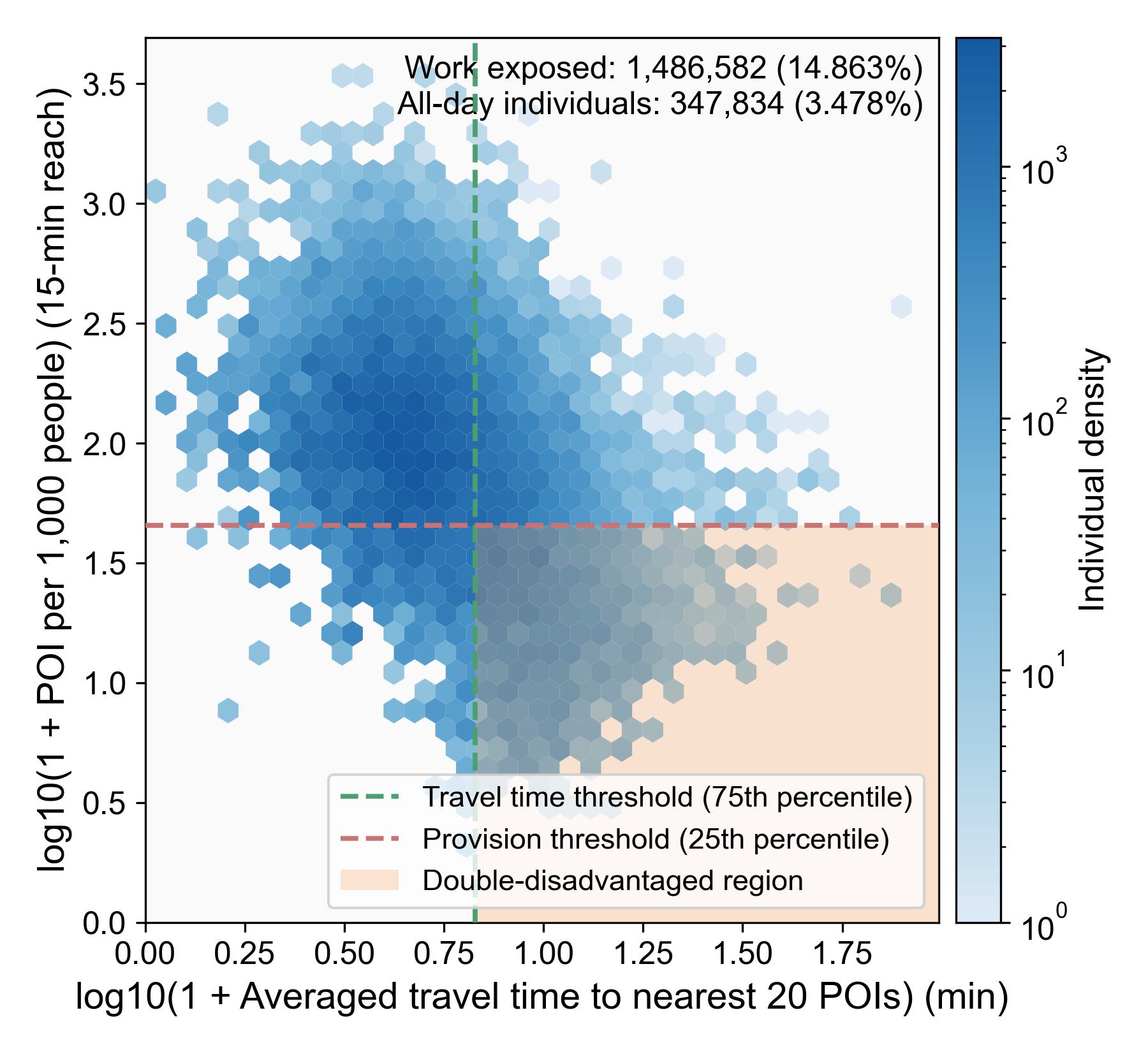}
\end{minipage}
\hspace{0.02\textwidth}
\begin{minipage}[t]{0.45\textwidth}
    \textbf{(b)}\\[2pt]
    \includegraphics[width=\linewidth]{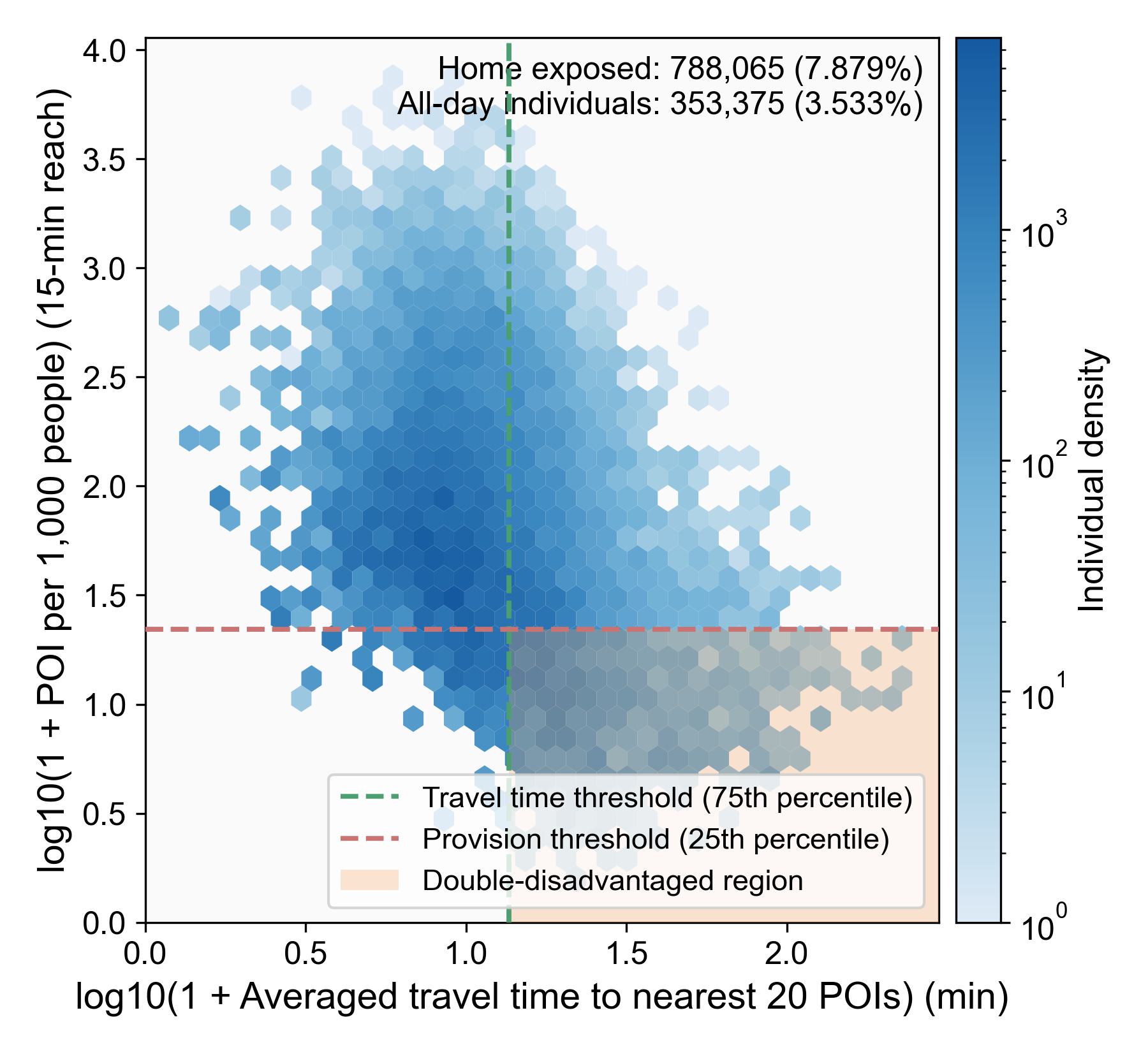}
\end{minipage}

\medskip

\begin{minipage}[t]{0.45\textwidth}
    \textbf{(c)}\\[2pt]
    \includegraphics[width=\linewidth]{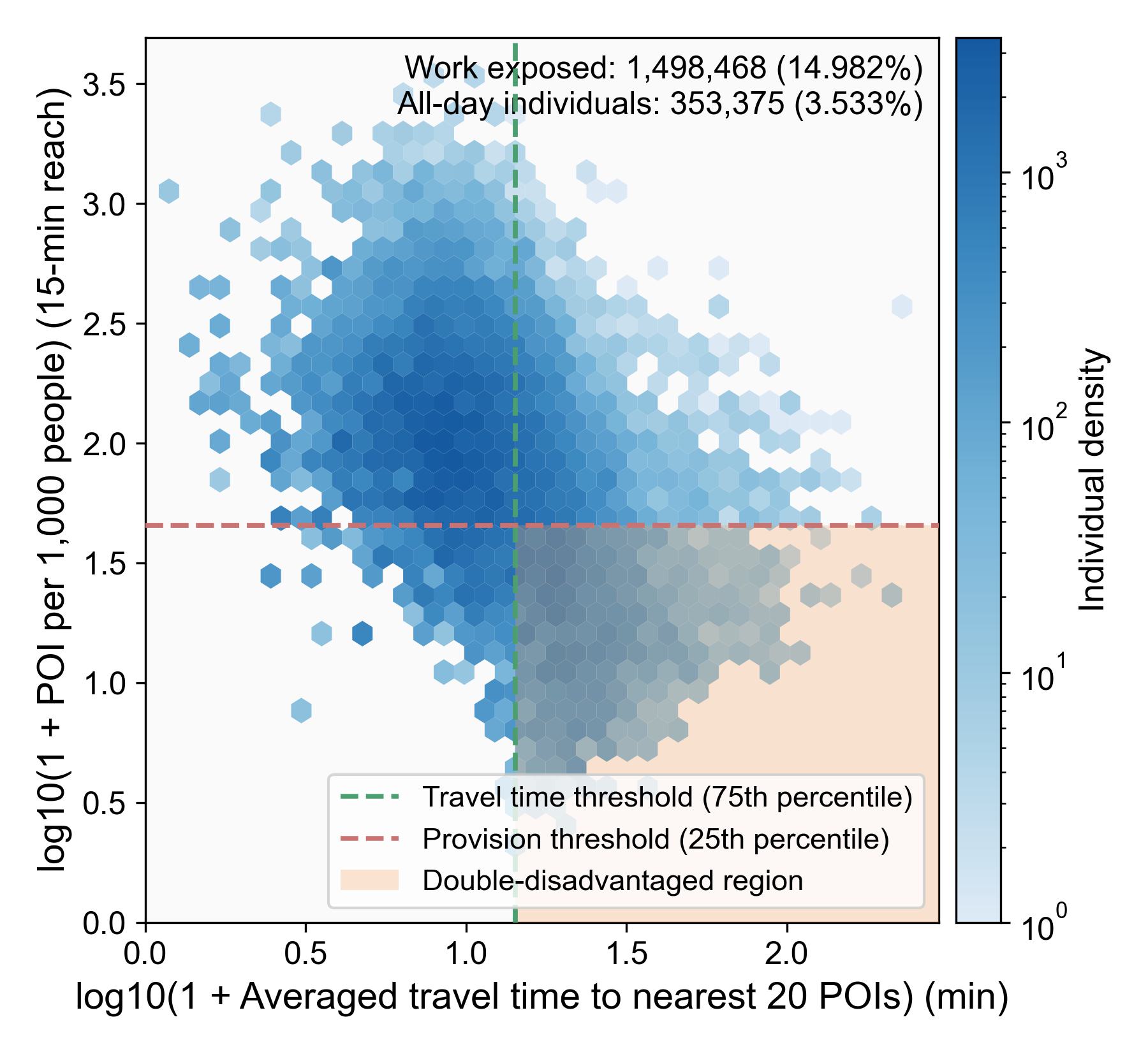}
\end{minipage}

\caption{Hexbin plots of proximity time versus per-capita provision for green spaces, showing transport mode--context combinations not included in Fig.~\ref{fig:double_disadvantage}. Each hexagonal bin is coloured by the number of individual observations; dashed lines indicate the quartile thresholds defining double disadvantage.
(a) Cycling -- workplace context.
(b) Walking -- residential context.
(c) Walking -- workplace context.
Statistics in the upper-right corner of each panel correspond to the respective rows in Table~\ref{tab:double_disadvantage_extended}.}
\label{fig:appendix_green_hexbin}
\end{figure}

\begin{figure}[H]
\centering
\begin{minipage}[t]{0.45\textwidth}
    \textbf{(a)}\\[2pt]
    \includegraphics[width=\linewidth]{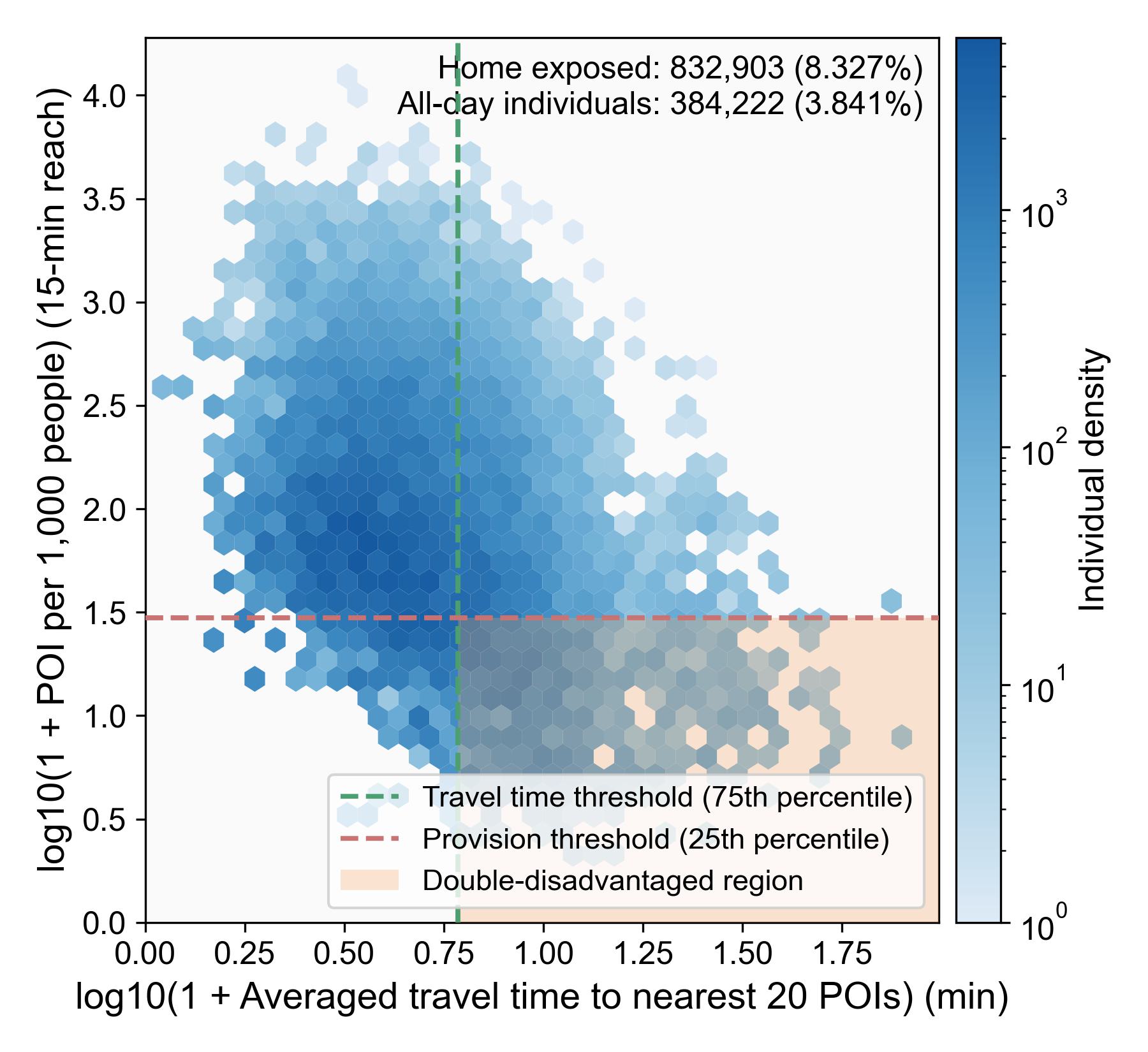}
\end{minipage}
\hspace{0.02\textwidth}
\begin{minipage}[t]{0.45\textwidth}
    \textbf{(b)}\\[2pt]
    \includegraphics[width=\linewidth]{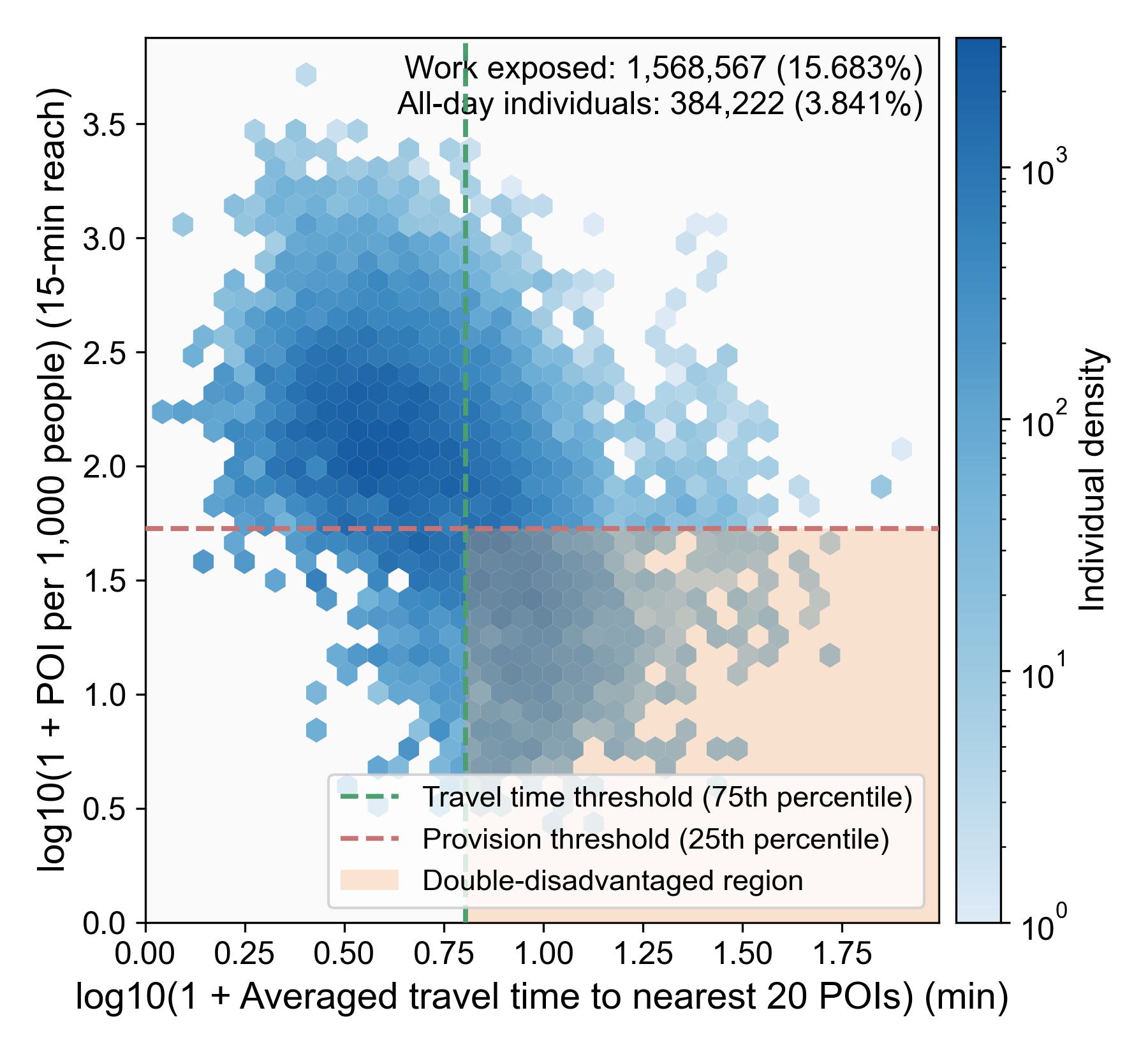}
\end{minipage}

\medskip

\begin{minipage}[t]{0.45\textwidth}
    \textbf{(c)}\\[2pt]
    \includegraphics[width=\linewidth]{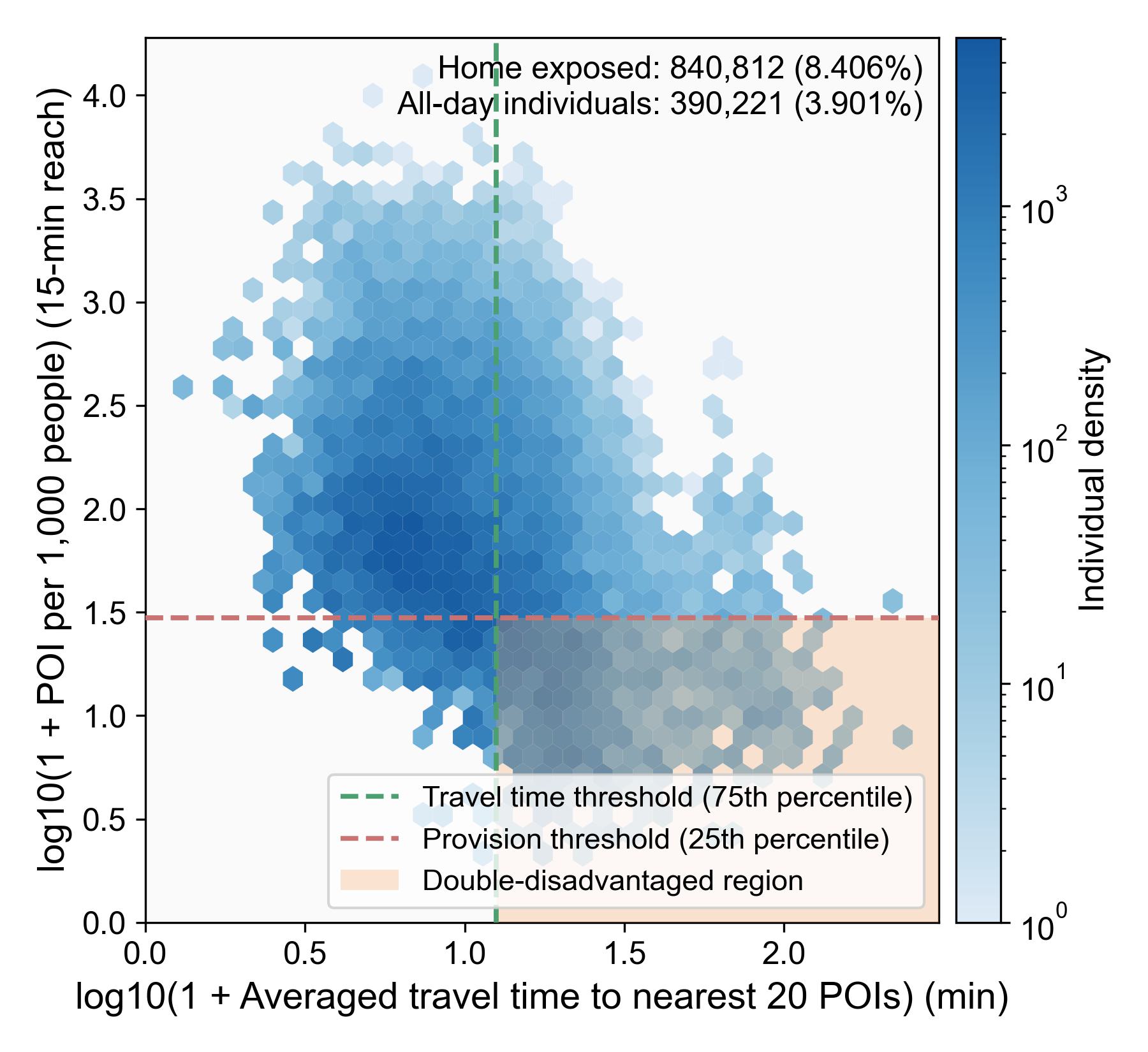}
\end{minipage}
\hspace{0.02\textwidth}
\begin{minipage}[t]{0.45\textwidth}
    \textbf{(d)}\\[2pt]
    \includegraphics[width=\linewidth]{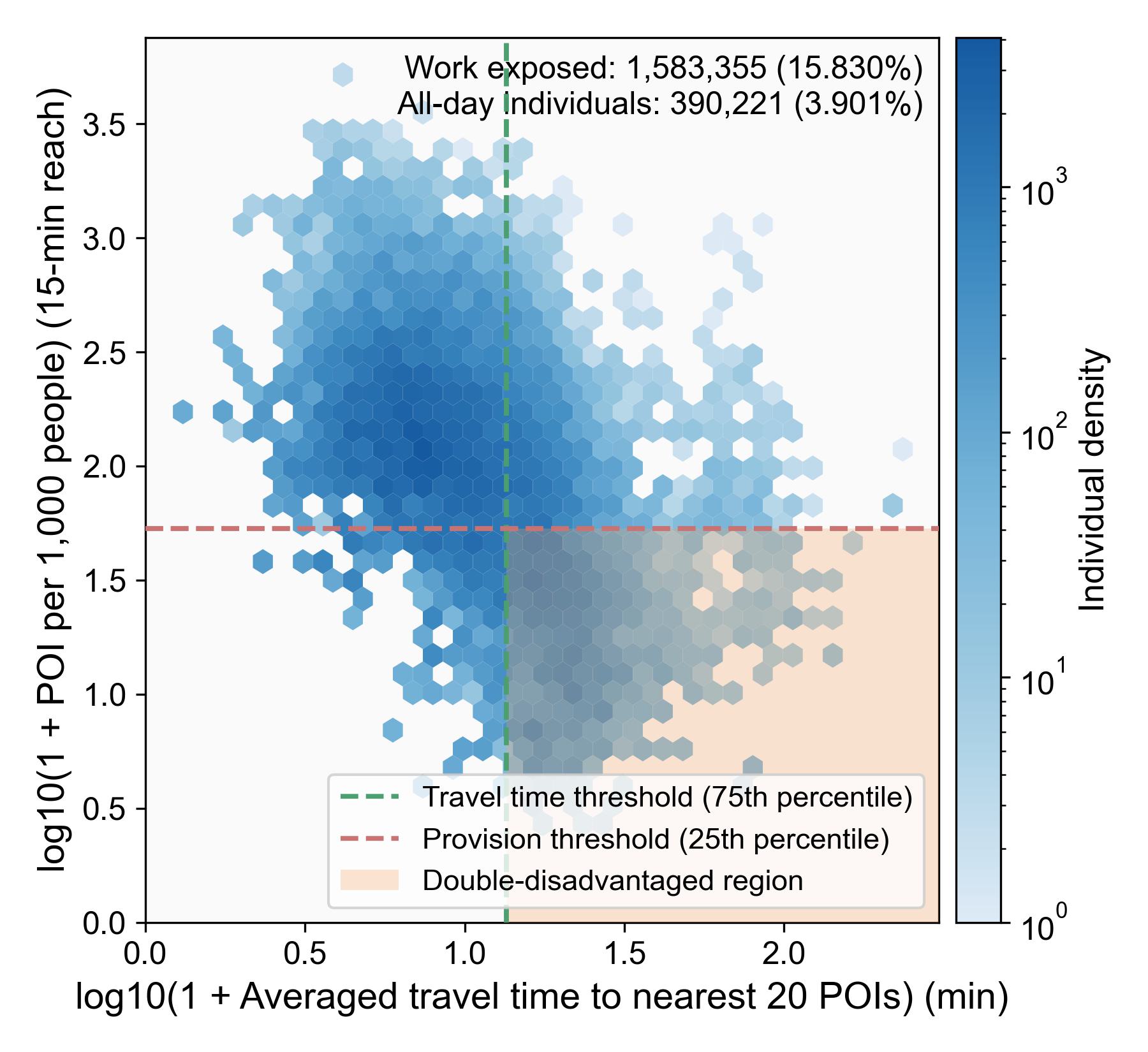}
\end{minipage}
\caption{Hexbin plots of proximity time versus per-capita provision for medical facilities under all transport mode--context combinations. Each hexagonal bin is coloured by the number of individual observations; dashed lines indicate the quartile thresholds defining double disadvantage.
(a) Cycling -- residential context (also shown as Fig.~\ref{fig:double_disadvantage}d in the main text).
(b) Cycling -- workplace context.
(c) Walking -- residential context.
(d) Walking -- workplace context.
Statistics in the upper-right corner of each panel correspond to the respective rows in Table~\ref{tab:double_disadvantage_extended}.}
\label{fig:appendix_medical_hexbin}
\end{figure}

\begin{figure}[H]
\centering
\begin{minipage}[t]{0.29\textwidth}
    \textbf{(a1)}\\[2pt]
    \includegraphics[width=\linewidth]{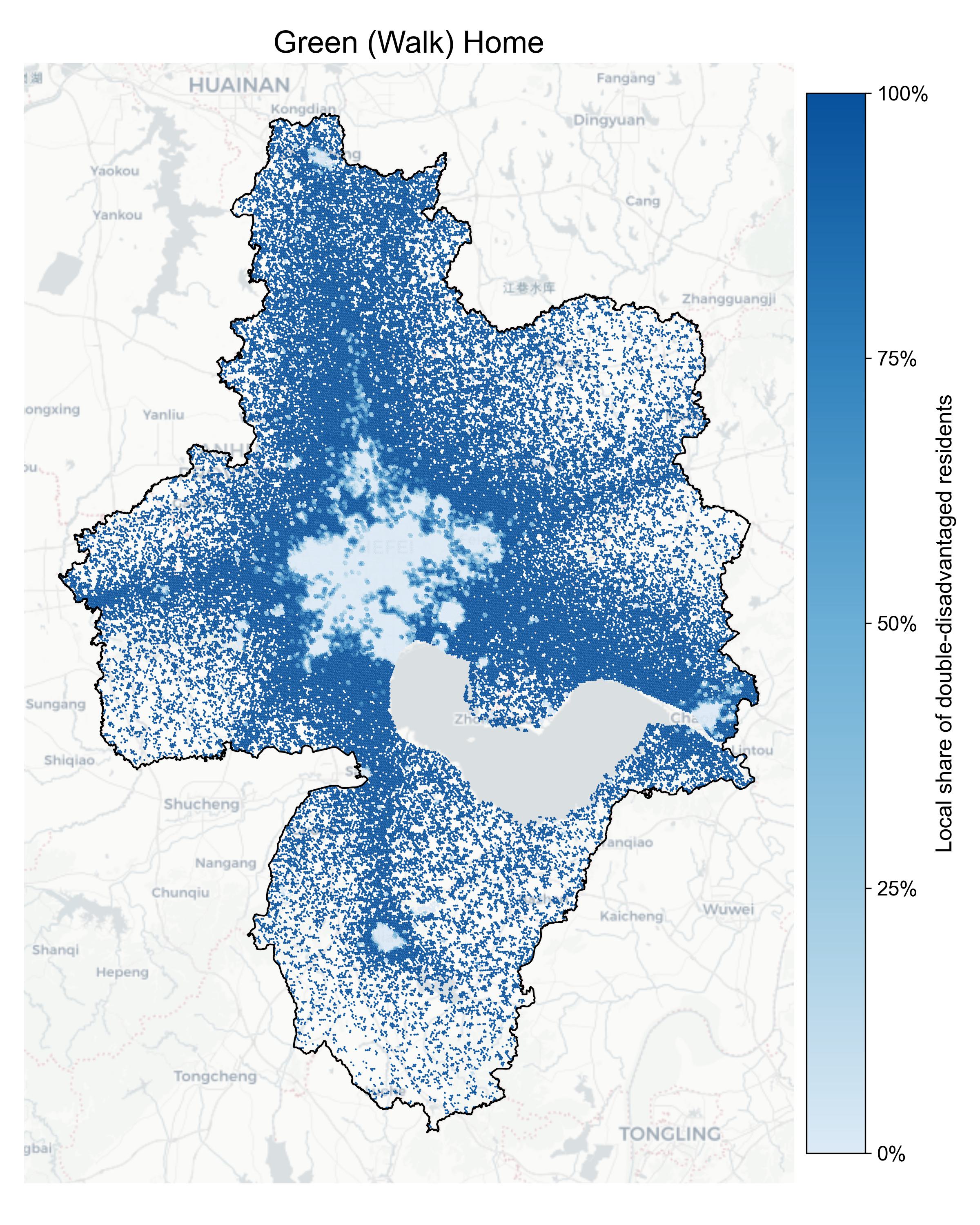}
\end{minipage}
\hspace{0.015\textwidth}
\begin{minipage}[t]{0.29\textwidth}
    \textbf{(b1)}\\[2pt]
    \includegraphics[width=\linewidth]{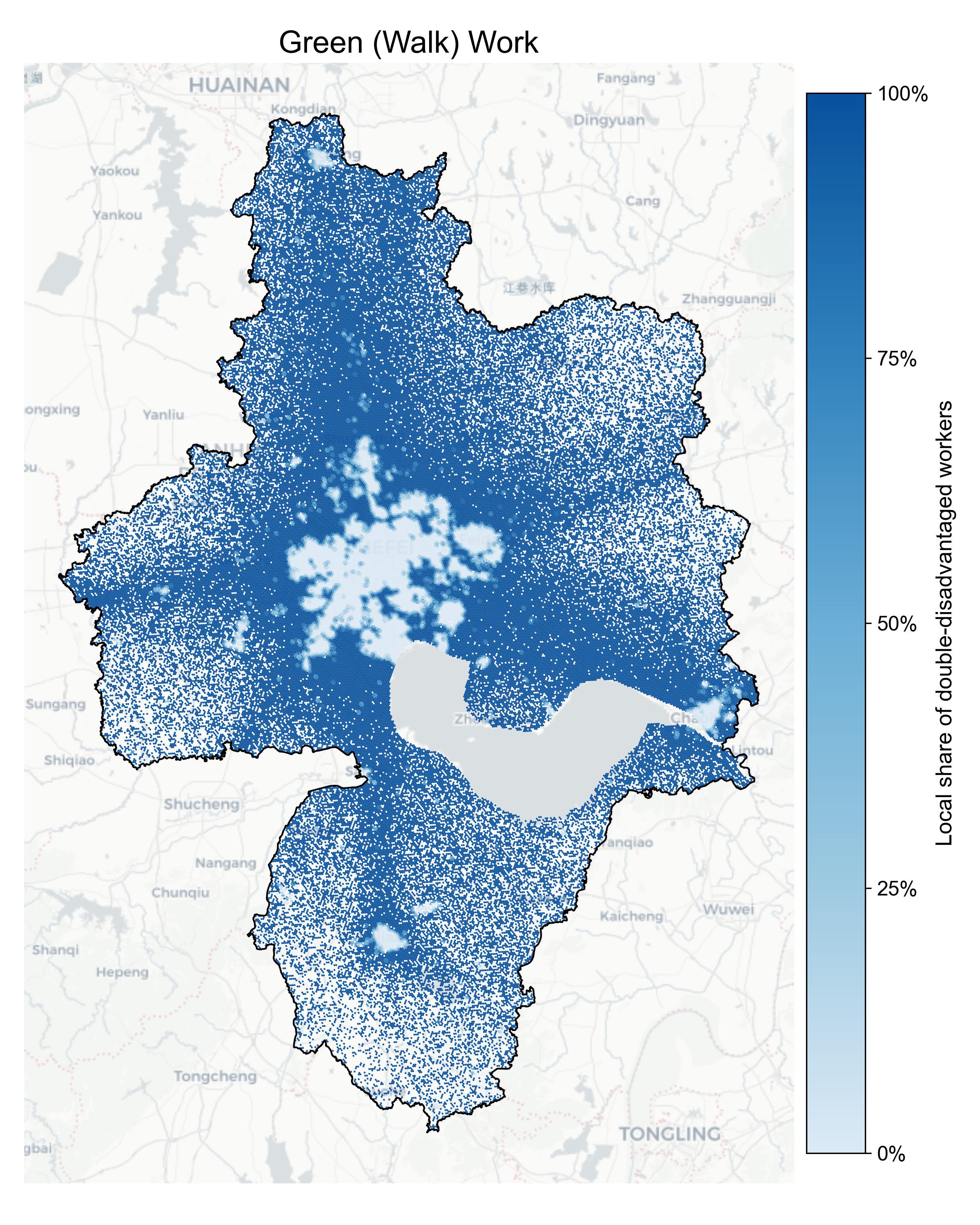}
\end{minipage}
\hspace{0.015\textwidth}
\begin{minipage}[t]{0.29\textwidth}
    \textbf{(c1)}\\[2pt]
    \includegraphics[width=\linewidth]{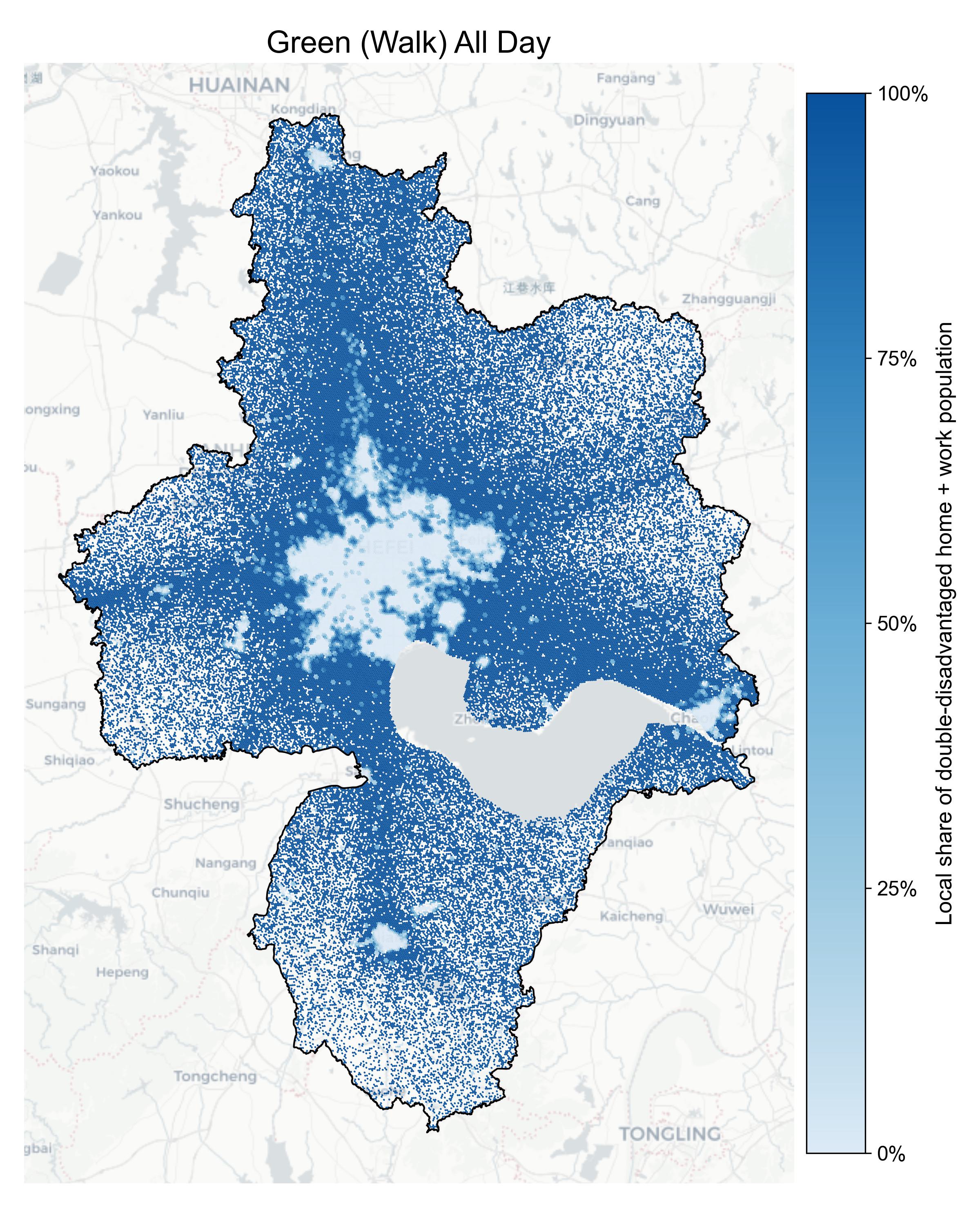}
\end{minipage}

\begin{minipage}[t]{0.29\textwidth}
    \textbf{(a2)}\\[2pt]
    \includegraphics[width=\linewidth]{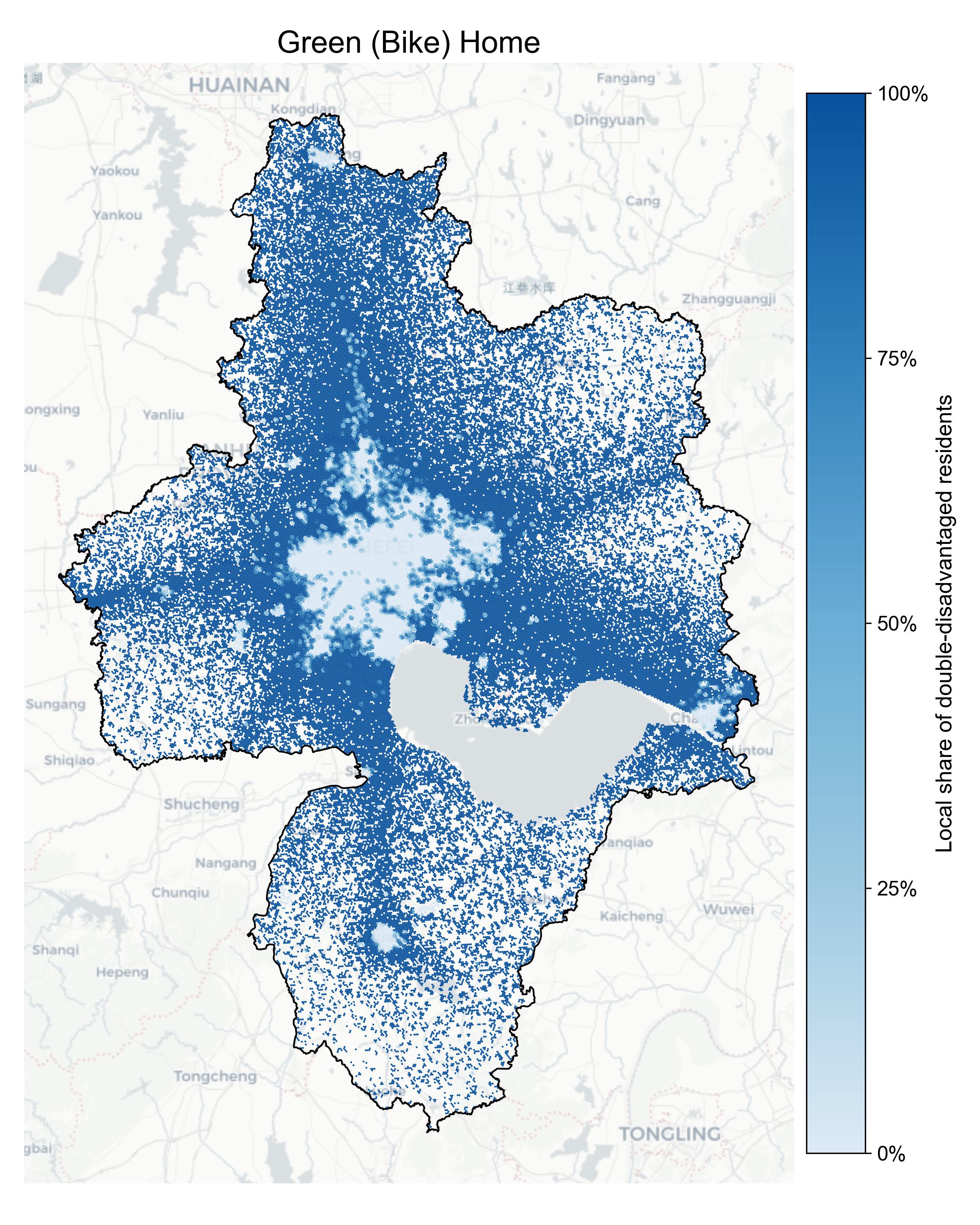}
\end{minipage}
\hspace{0.015\textwidth}
\begin{minipage}[t]{0.29\textwidth}
    \textbf{(b2)}\\[2pt]
    \includegraphics[width=\linewidth]{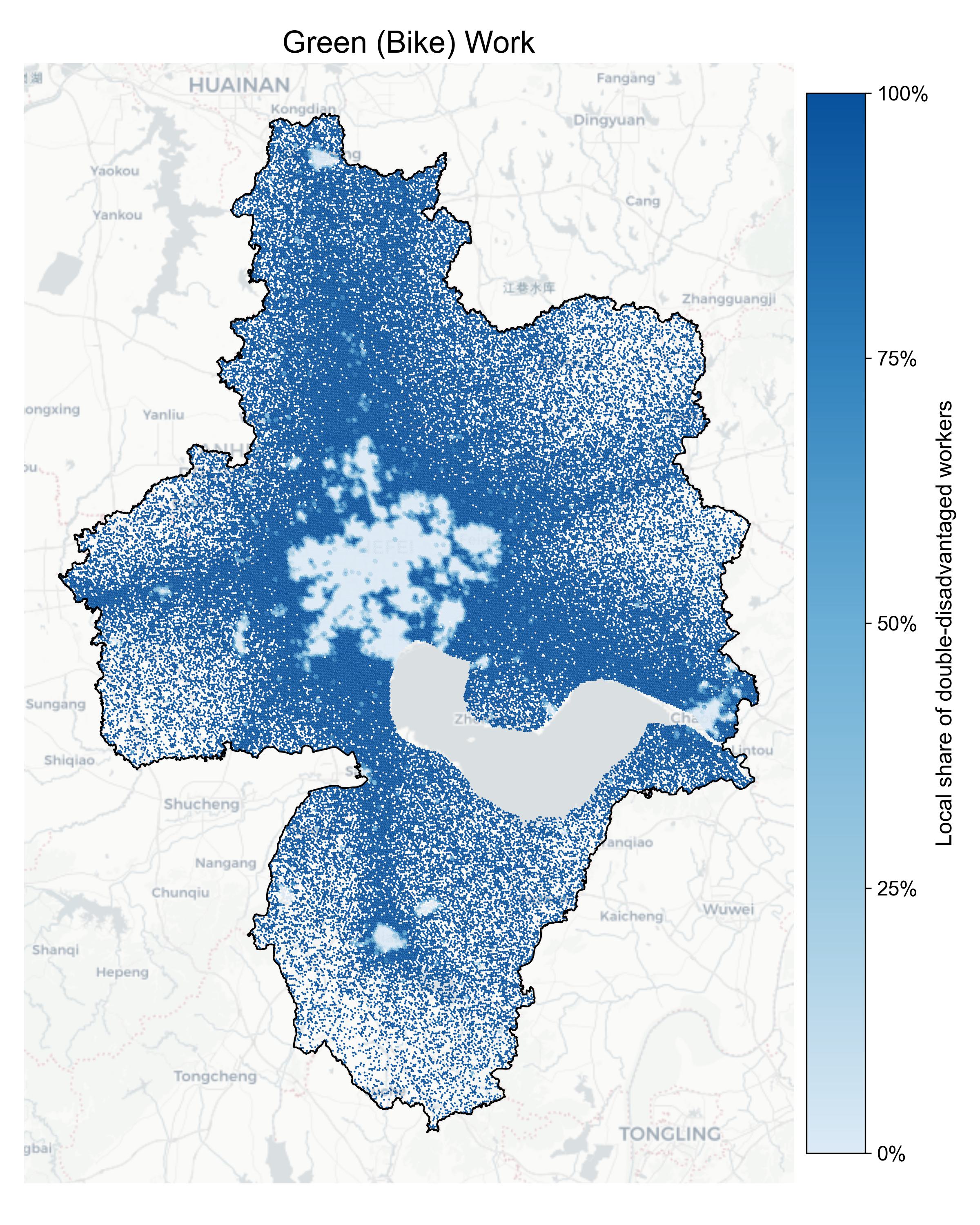}
\end{minipage}
\caption{Double-disadvantaged areas in access to green spaces under walking and cycling conditions, expressed as the share of disadvantaged individuals relative to the corresponding reference population. (a) Residential exposure, normalised by residents. (b) Workplace exposure, normalised by workers. (c) All-day exposure, normalised by the combined residential-workplace reference population. (1) Walking condition. (2) Cycling condition.}
\label{fig:appendix_green_walk}
\end{figure}

\clearpage 
\bibliographystyle{cas-model2-names}

\bibliography{reference}

\bio{}
\endbio
\printcredits
\end{document}